\documentclass[journal]{IEEEtran}
\usepackage{ graphicx}
\usepackage[utf8]{inputenc}
\usepackage[english]{babel}
\usepackage{mathtools}
\usepackage{multirow}
\usepackage{multicol}
\usepackage{longtable}
\usepackage{amsmath}
\usepackage[boxed,vlined]{algorithm2e}
\usepackage{hyphenat}
\usepackage{booktabs}
\usepackage[table,xcdraw]{xcolor}
\usepackage{cite}
\usepackage{subcaption}
\usepackage{skull}
\usepackage{pgfplots}
\usepackage{tikz}

\ifCLASSINFOpdf
\else
\fi
\begin{document}
	%
	\title{Encrypt Flip-Flop: A Novel Logic Encryption Technique For Sequential Circuits}
	%
	%
	%
	
	\author{Rajit Karmakar,~\IEEEmembership{Student Member,~IEEE,}
		Santanu Chattopadhyay,~\IEEEmembership{Senior Member,~IEEE,}
		\\ and Rohit Kapur,~\IEEEmembership{Fellow,~IEEE}\vspace{-0.5cm}
		\thanks{Rajit Karmakar and Santanu Chattopadhyay are with the Department
			of Electronics and Electrical Communication Engineering, Indian Institute of Technology, Kharagpur,
			India, 721032 e-mail: \{rajit,santanu\}@ece.iitkgp.ernet.in}
		\thanks{Rohit Kapur is with Synopsys, USA, e-mail: Rohit.Kapur@synopsys.com}
		\thanks{This work is partially supported by the research project entitled “Synopsys CAD Laboratory Projects (CADL)“, sponsored by Synopsys Inc., USA.}}
	\maketitle

	\begin{abstract}
		Logic Encryption is one of the most popular hardware security techniques which can prevent IP piracy and illegal IC overproduction. It introduces obfuscation by inserting some extra hardware into a design to hide its functionality from unauthorized users. Correct functionality of an encrypted design depends upon the application of correct keys, shared only with the authorized users. In the recent past, extensive efforts have been devoted in extracting the secret key of an encrypted design. At the same time, several countermeasures have also been proposed by the research community to thwart different state-of-the-art attacks on logic encryption. However, most of the proposed countermeasures fail to prevent the powerful SAT attack. Although a few researchers have proposed different solutions to withstand SAT attack, those solutions suffer from several drawbacks such as high design overheads, low output corruptibility, and vulnerability against removal attack. Almost all the known logic encryption strategies are vulnerable to scan based attack. In this paper, we propose a novel encryption technique called \textit{Encrypt Flip-Flop}, which encrypts the outputs of selected flip-flops by inserting multiplexers (MUX). The proposed strategy can thwart all the known attacks including SAT and scan based attacks. The scheme has low design overhead and implementation complexity. Experimental results on several ISCAS'89 and ITC'99 benchmarks show that our proposed method can produce reasonable output corruption for wrong keys.

	\end{abstract}
	

	%
	\IEEEpeerreviewmaketitle

	\section{Introduction}
	Ever increasing market demand for smarter, faster and smaller products motivates the electronics design industry to develop complex chips with a wide range of functionalities like digital, analog, radio frequency, photonic, integrated into a single chip. Manufacturing these complex chips requires advanced mixed technology fabrication facilities. The enormous cost of setting up and maintaining such fabrication lab (cost of owning a foundry is about \$5 billion \cite{online1}) is the main impediment for small design houses to own an in-house foundry. However, globalization in the semiconductor industry has facilitated integrated circuit (IC) designers to outsource the fabrication of their designs to offshore foundries. Although this trend significantly cuts down the cost, at the same time, it has also opened the backdoor for several security vulnerabilities like Intellectual Property (IP) piracy, counterfeiting, reverse engineering, overbuilding, insertion of hardware Trojans \cite{rostami2014primer,roy2008epic}. The accessibility of the GDS-II file to the third party foundry personnel exposes the IP of a design. An untrustworthy user in the foundry may reverse engineer the GDS-II file and claim the ownership of the IP. Illegal overproduction and selling the excess ICs is another possible trend of stealing a design. These kinds of design thefts cost the semiconductor industry a loss of several billions of dollars, every year \cite{online2}. To withstand these security threats, Design-for-Security (DfS) has emerged to be a conjoined part of IC design.
	
\begin{figure}[t]
		\centering
		\includegraphics[scale = 0.4]{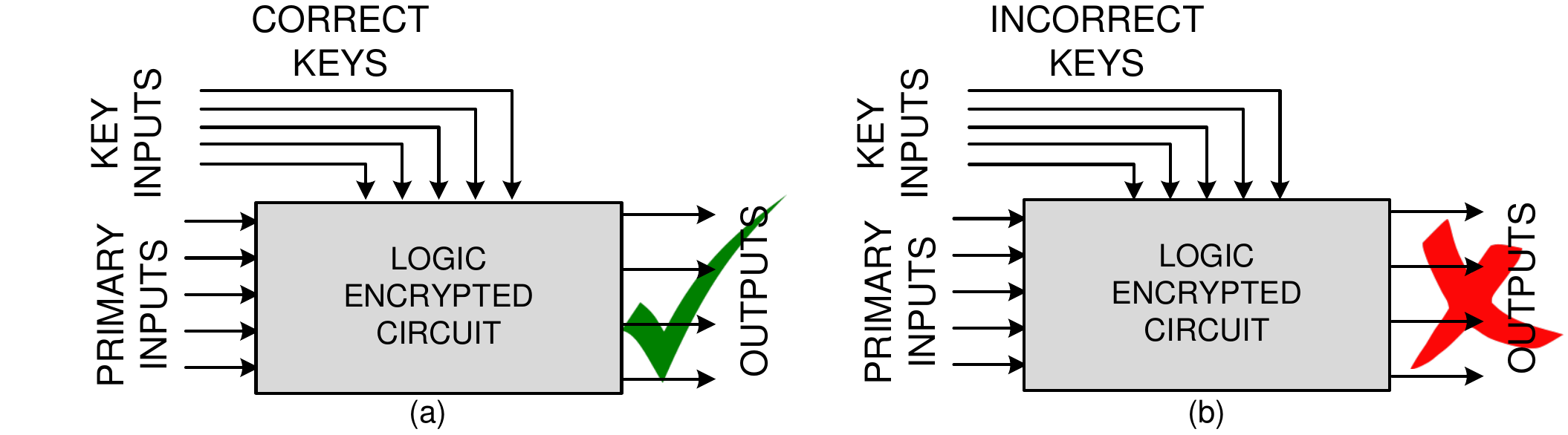}
		\vspace{-0.3cm}
		\caption{Block diagram of Logic Encryption \cite{karmakar2017new}}
		\label{fig: basic_block_diagram}
		\vspace{-0.7cm}
	\end{figure}

	Logic encryption is a popular countermeasure to restrict IP piracy and illegal overproduction by the foundry. Using logic encryption, a designer can introduce some redundant logic elements (key-gates) into a circuit to conceal its functionality from a third party foundry. Correct functionality of an encrypted IC depends on the application of the correct keys to the key-gates. The fabricated IC is activated by applying the secret keys when it returns back to the design house from the foundry \cite{yasin2016activation}. These secret keys are stored in a tamper-proof memory inside the chip. Unavailability of the correct keys inhibits an unauthorized user from reverse engineering the GDS-II file, and claiming the ownership of the design. Illegally over-produced ICs cannot be sold in the market as these chips do not exhibit correct functionality until they are activated with the exact keys. Figure \ref{fig: basic_block_diagram} shows a basic block diagram of logic encryption.
	
	\begin{figure*}[t]
		\centering
		\includegraphics[scale = 0.18]{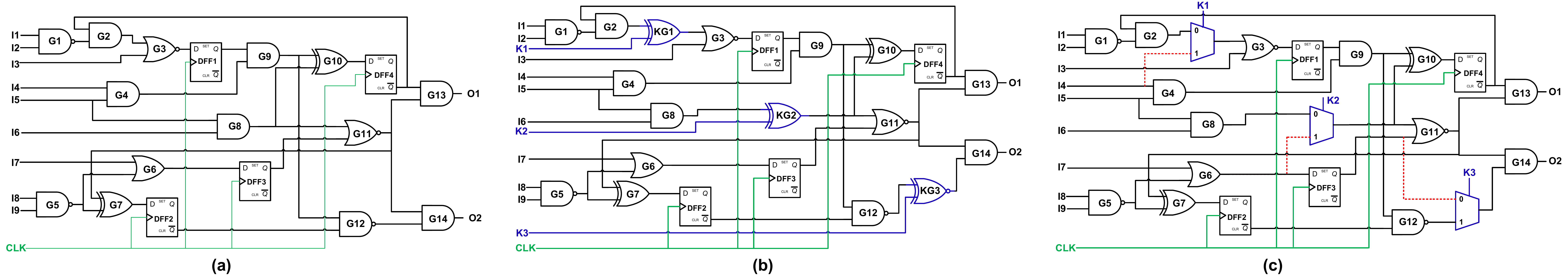}
		\vspace{-0.3cm}
		\caption{An example of logic encryption. (a) original netlist, (b) encryption using XOR/XNOR gates, (c) encryption using MUX}
		\label{fig:example}
		\vspace{-0.7cm}
	\end{figure*}
	
	\section{Background and Preliminary Ideas} 
	Logic encryption can be either sequential or combinational in nature. Sequential logic encryption \cite{chakraborty2009harpoon} introduces a Finite State Machine (FSM) which uses some of the primary inputs of the original circuit as its inputs. The state transition graph is modified with some additional logic states, called black states. A correct input sequence is required to reach a valid state, which allows the correct functionality of the encrypted circuit. Wrong input sequence restricts the operation of the chip by entering into one of the black states. On the other hand, combinational logic encryption techniques use XOR/XNOR gates \cite{roy2008epic, rajendran2015fault, yasin2015improving} to encrypt a netlist. Few other methods use AND/OR gates \cite{dupuis2014novel}, multiplexers (MUX) \cite{rajendran2015fault, zhang2016practical},  Look Up Tables (LUT) \cite{baumgarten2010preventing} as key-gates. The XOR/XNOR based encryption technique was introduced in EPIC \cite{roy2008epic}. This method randomly inserts XOR/XNOR gates (key-gates) into a design. One input of the XOR/XNOR gate is connected to some internal line of the circuit, while the other input serves as a key-input. Figure \ref{fig:example}(b) shows a typical example of XOR/XNOR based logic encryption which encrypts the circuit of Figure \ref{fig:example}(a) using three key-gates, KG1, KG2, and KG3. These key-gates are configured as buffers upon applying correct keys, else they invert the lines, leading to wrong output for invalid keys.
	
	In MUX-based encryption \cite{rajendran2015fault}, several 2 input-MUXes are inserted into a design. Two different lines (one true line and one false line) of the circuit are connected to the inputs of a MUX, while the select line of the MUX acts as the key. Correct values of the keys propagate the values at the true lines to the outputs of the MUXes. Figure \ref{fig:example}(c) shows an example of MUX based logic encryption. Another MUX-based encryption considers obfuscation cell (OC) (a combination of a MUX and an inverter) to encrypt a design \cite{zhang2016practical}. In \cite{wang2016secure}, Wang $et$ $al.$ have proposed an encryption technique which uses a combination of MUX and camouflage connectors as configurable logic units to replace certain logic gates. 
	
	To activate an encrypted IC, an unauthorized user must extract the correct keys. If an IC with $M$ inputs is encrypted with $K$ keys, a brute force attack requires $2^M$ observations from an active IC and O($2^{M+K}$) computations on an encrypted design. This is practically impossible for sufficiently large values of $M$ and $K$. Although these methods seem to secure a design from theft, some later research works have pointed out several shortcomings, which can expose a logically encrypted circuit to several attacks. However, every time a new attack has been proposed, researchers have come forward with a modified strategy to logically encrypt a design, which eventually has evolved logic encryption techniques towards more secure and better ways to thwart different security vulnerabilities.

	\section{Evolution Of Logic Encryption}\label{sec:evo}
	One important criterion of logic encryption is that incorrect keys must produce wrong outputs. This implies that the effect of incorrect keys should propagate to the outputs and corrupt some of the output bits. The EPIC method \cite{roy2008epic} of logic encryption randomly inserts key-gates into the design. However, random insertion of key-gates does not always ensure high output corruption for incorrect keys, as the effect of wrong key may get masked by other inputs, thus, may not propagate to the output. To ensure high output corruption for invalid keys, a fault analysis based key-gate location selection approach has been proposed in \cite{rajendran2015fault}.
	
	\textbf{Fault analysis based logic encryption \cite{rajendran2015fault}:} This method uses three basic phenomena, fault excitation, fault propagation and fault masking of IC testing to identify several locations in the circuit, where if any fault occurs (either s-a-0 or s-a-1), propagates to the output and corrupts a maximum number of output bits for most of the applied input patterns. Insertion of key-gates in these locations ensures high output corruption for wrong keys.
	
	Although, the fault analysis based technique \cite{rajendran2015fault} fulfills the criterion of high output corruption for wrong keys, some later research works have shown that both random and fault analysis based key-gate insertion approaches are vulnerable to several attacks, such as, logic cone analysis \cite{lee2015improving}, hill-climbing \cite{plaza2015solving}, path sensitization \cite{yasin2015improving}, and SAT-based \cite{subramanyan2015evaluating} attacks.
	
	The basic requirements for these attacks are 1) an encrypted netlist, available in the foundry, and 2) a functional IC, available in the market. An attacker applies the same input to both the circuits and compares the outputs to extract the keys. Using these attacks an attacker can extract the keys of an encrypted design. 
	
	\textbf{Logic cone analysis based attack \cite{lee2015improving}:} This attack aims to minimize the effort of brute force attack by following a divide and conquer strategy to explore the keys. It checks the number of key-gates affecting each of the outputs and targets the output with the smallest number of key-gates in its input cone of dependency. A brute force attack on an output which is affected by the fewest number of key-gates is a feasible solution to extract a subpart of the keys. In each iteration, the process searches for the outputs with less number of key-gates in their input cone of dependencies and applies brute force to extract a small portion of the entire key set. To prevent logic cone analysis based attack, Lee $et$ $al.$ have proposed strategic insertion of some MUXes into the design \cite{lee2015improving}. The process creates more overlap between the logic cones, which increases the number of key-gates in the input cone of dependency of each output.
	
	\textbf{Hill Climbing Attack \cite{plaza2015solving}:} This attack starts by applying an initial random key to an encrypted netlist and measuring the Hamming distance between the obtained and the expected correct outputs of a functional chip. With the ultimate goal of obtaining zero Hamming distance for any set of input patterns, each iteration of the attack takes a decision whether to flip a key-bit or not. The attack succeeds on the finding of such a key which can produce zero Hamming distance between the observed and the correct outputs for all input patterns.

	\textbf{Path sensitization attack \cite{yasin2015improving}:} A key-bit can be sensitized to the output by selecting specific input pattern if no other key-gates interfere in the sensitization path of the key-bit. A similar kind of input pattern is applied to both functional IC and encrypted netlist. Observation of the outputs of these two circuits can reveal the key values.

	\textbf{Strong logic locking \cite{yasin2015improving}:} To prevent path sensitization attack, Yasin $et$ $al.$ have proposed a strong logic locking strategy \cite{yasin2015improving} by inserting key-gates in such locations which forms a clique where all nodes (key-gates) interfere with each other. The size of the clique reflects the length of the key. This strategy ensures that sensitization of any key-gate to an output requires applying suitable values to the primary inputs as well as other key inputs. As the other keys are not known to an attacker, no key can be sensitized to the output. One drawback of this topology dependent key-gate insertion strategy is that it may not offer ample key-gate locations to encrypt a design with sufficient number of keys. Moreover, as the key-gates are placed with an objective to increase the clique size, it does not always ensure high output corruption for wrong keys.
	
	\textbf{External key-dependency \cite{karmakar2017new, karmakar2017enhancing}:} To overcome the drawbacks of strong logic locking, Karmakar $et$ $al.$ have proposed an iterative approach which can prevent path sensitization attack as well as ensure high output corruption for invalid keys \cite{karmakar2017new}. The nonlinear interdependency among the primary and secondary keys of this external key-dependency based encryption strategy helps to thwart both hill climbing and logic cone based attacks. However, these methods \cite{karmakar2017new, karmakar2017enhancing} incur some extra hardware to incorporate the key-dependency unit into the design, which also increases the power and delay overheads of the design.
	
	\textbf{SAT-based attack \cite{subramanyan2015evaluating}:}  Recently, a powerful SAT attack was proposed in \cite{subramanyan2015evaluating}. This attack uses a SAT-based algorithm to extract the keys of a logically encrypted combinational circuit. The attack algorithm iteratively searches for a special set of $distinguished$ $input$ $patterns$ ($DIPs$), which help to reduce the key search space by eliminating the incorrect keys. A $DIP$ ensures that at least two different keys produce different outputs. Comparison of the outputs with the output of a functional chip, for the same $DIP$, helps to eliminate at least one or both the keys as incorrect keys. The attack shows that using a limited number of $DIPs$, all the incorrect keys can be eliminated and an equivalent set of correct keys can be revealed. Another SAT-based attack called AppSAT was proposed in \cite{shamsi2017appsat}, which can approximately deobfuscate an encrypted netlist with very low error rate.  
	
	\textbf{SAT-resilient techniques:} The complexity of SAT-based attack depends on the complexity of the circuit as well as the number of $DIP$s required to eliminate all the wrong keys. To prevent SAT attack, Yasin $et$ $al.$ have proposed to integrate some extra hardware (called SARLock) \cite{yasin2016sarlock} with strong logic locking \cite{yasin2015improving}, that increases the effort of SAT attack by exponentially increasing the number of $DIPs$ to eliminate all the incorrect keys. The proposed SARLock method modifies the outputs in such a way that an incorrect key produces a wrong output only for a specific input pattern. Therefore, a $DIP$ can eliminate only one incorrect key. For a sufficiently long key, the exponential number of required $DIP$s makes the SAT attack impossible. However, some later research \cite{yasin2017lock} has shown that SARLock is vulnerable to removal attack. To mitigate the removal attack, Yasin $et$ $al.$ have proposed a new SAT-resilient encryption technique called TTLock \cite{yasin2017lock} which modifies a logic cone by flipping the output for a secret input pattern and restores the flip for correct keys. Another SAT-based attack called Double DIP \cite{shen2017double} has been proposed in recent time, which can avoid the exponential iteration of key search process incorporated by the SARLock method. Yang $et$ $al.$ have also proposed to use an Anti-SAT block \cite{xiemitigating} to exponentially increase the number of SAT attack iterations to reveal the correct key. However, the security of Anti-SAT block can be bypassed using a Signal Probability Skew (SPS) attack \cite{yasinsecurity}. In \cite{xu2017novel}, Xu $et$ $al.$ have also shown that both SARLock and Anti-SAT are vulnerable to a new bypass attack. In \cite{xie2017delay}, Xie $et$ $al.$ have proposed to use tunable delay key-gates (TDK) to encrypt a design. The proposed Delay Locking strategy considers two keys for each TDK, one for functional locking and the other for manipulating the delay. The introduction of timing violation for wrong delay keys helps this method to thwart SAT attack. Another SAT-resilient secure cell design technique has been proposed in \cite{guin2017novel}. A cyclic obfuscation based SAT-resilience encryption technique has been proposed in \cite{shamsi2017cyclic}, which creates logical loop in a circuit by adding dummy wires and gates. The approach ensures that all the inserted cycles have multiple ways to open. As the circuit can no longer be represented as a directed acyclic graph (DAG), the conventional SAT-based attack cannot be applied to extract the keys. However, Zhou $et$ $al.$ recently proposed an algorithm called CycSAT \cite{zhoucycsat}, which can effectively decrypt cyclic obfuscation.
	
	In recent times, several transistor level logic encryption techniques have been proposed in \cite{juretus2016reduced,bi2016enhancing}. Chen $et$ $al.$ have proposed a low overhead gate replacement technique \cite{chen2017low} for logic encryption. The proposed technique can significantly reduce the area, power, and delay overheads, compared to the typical XOR/XNOR based encryption, however, it fails to thwart SAT attack.

	\section{Motivation And Contribution Of The Paper}
	In the previous section, we have observed several shortcomings of different logic encryption strategies. For example, strong logic locking \cite{yasin2015improving} can thwart path sensitization and hill climbing attacks, however, sometimes fails to encrypt a design with a sufficiently large number of keys. Similarly, external key-dependency based approach \cite{karmakar2017new} prevents path sensitization, hill climbing, and logic cone based attacks at the cost of higher hardware, power, and delay overheads, compared to other methods. We have also observed that most of the logic encryption strategies are vulnerable to SAT-based attack. Although the SARLock \cite{yasin2016sarlock}  and Anti-SAT \cite{xiemitigating} methods restrict the SAT attack, they require extra hardware to increase the effort of SAT-attack. The SARLock method modifies a design in such a way that even for a wrong key, the circuit produces correct output for most of the inputs, which may not be a desirable criterion from a designer's point of view. Moreover, both SARLcok and Anti-SAT methods are vulnerable to removal attack. Most of the existing logic encryption strategies fail to prevent all the known state-of-the-art attacks while simultaneously fulfilling the basic requirements like high output corruptibility, low design overhead and low implementation complexity. These observations clearly show that despite a substantial amount of research, logic encryption methods are yet to get matured enough, which leaves room for further improvements. The demand for a low overhead and secured way to logically encrypt a design has motivated us to develop a new logic encryption strategy.
	
	The main contributions of the paper are as follows.
	\begin{enumerate}
		\item We propose a scan based attack which exploits the DfT infrastructure to partition a circuit into multiple smaller sub-circuits and attack them individually. The attack can drastically reduce the complexity of any state-of-the-art attack on any logically encrypted sequential circuit.    
		\item To prevent the proposed attack, we introduce a new logic encryption strategy, which encrypts the outputs of the flip-flops. The proposed \textit{Encrypt Flip-flop} strategy ensures that the scan chains do not leak any key information, thus, prevent scan based attack.
		\item The proposed encryption strategy restricts the controllability and observability of the flip-flops of the scan chains. This inhibits an attacker to apply SAT-based attack by converting a sequential design to a combinational one using scan facility. Unlike other methods our method does not incur any extra hardware to prevent SAT attack.
		\item The proposed low overhead encryption strategy can also prevent other state-of-the-art attacks, like path sensitization, hill climbing, and logic cone based attacks.
	\end{enumerate}

	\section{Proposed Scan Based Attack On Logic Encryption}\label{leb:scan_attack}
	In this section, we present an attack on conventional logic encryption techniques, which uses the phenomenon of scan based side channel attack. We show that the scan chains of a design can be exploited to drastically reduce the complexity of several state-of-the-art attacks on logic encryption.
	
\begin{figure}[!ht]
		\centering
		\includegraphics[scale = 0.2]{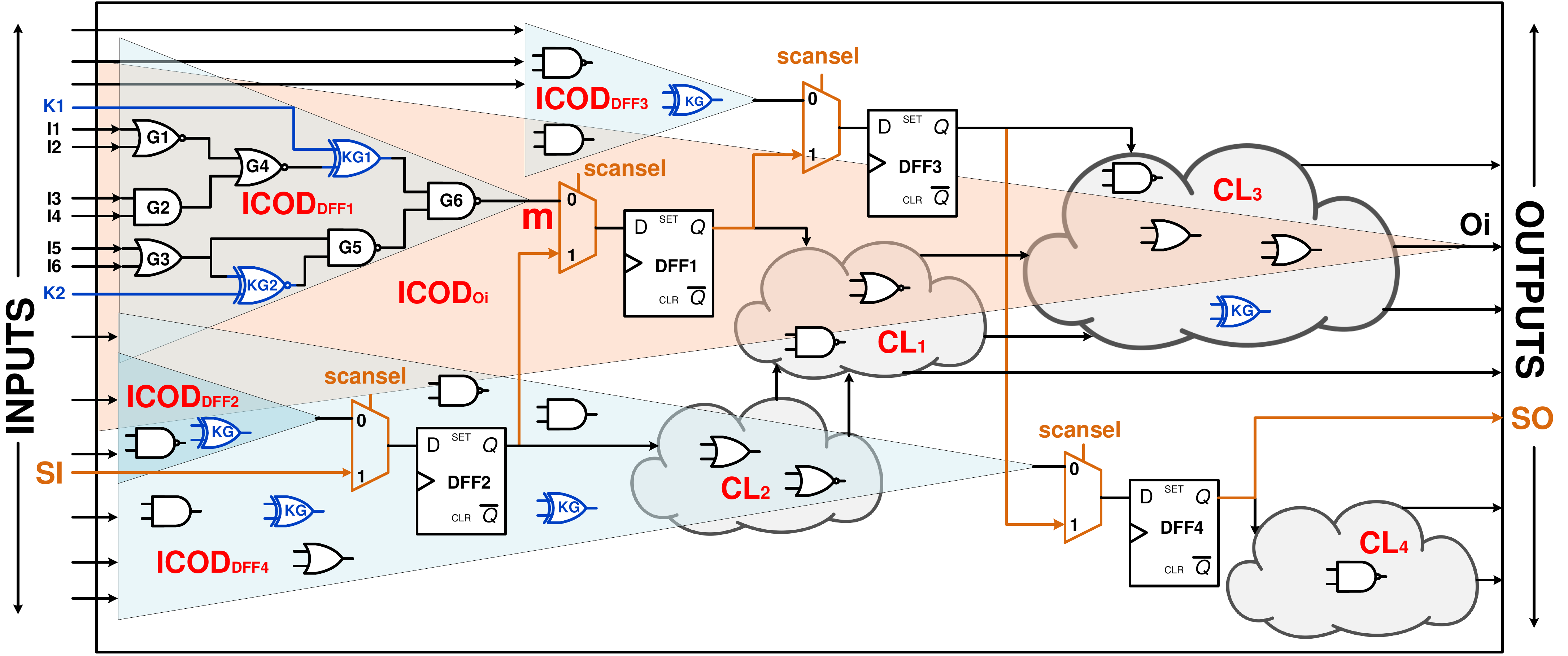}
		\caption{Basic idea of scan based attack on logic encryption}
		\label{fig:our_attack}
	\end{figure}

	\textbf{Attack Infrastructure:} Like other attacks on logic encryption, our proposed attack also requires an encrypted netlist and an activated IC. The attack assumes that the design contains flip-flops and the activated IC has DfT infrastructure \cite{cui2015ultra, ye2015diagnosis} (i.e. in full scan environment, all flip-flops get replaced by scan flip-flops, and are connected in a chain) for the purpose of infield testing and debugging.  Figure ~\ref{fig:our_attack} shows the block diagram of a circuit which contains four flip-flops connected in a scan chain. The attack assumes that an attacker can switch between functional and scan modes at any point of time.  
	
	\textbf{Attack Scenario:} We explain the attack using the example shown in Figure ~\ref{fig:our_attack}. Please note that in functional mode, the flip-flop $DFF1$ gets affected only by its input cone of dependency $ICOD_{DFF1}$. $ICOD_{DFF1}$ has six primary inputs ($I1 \rightarrow I6$), feeding several combinational logic gates. It also contains two key-gates with key-inputs $K1$ and $K2$. In functional mode, $DFF1$ stores the output of $ICOD_{DFF1}$ (line $m$). Similarly, the outputs of the $ICOD$ of other flip-flops also get stored in the corresponding flip-flops. We can observe the content of these flip-flops by switching from functional to scan mode and shifting them out through the scan out port $SO$. The observability introduced by the scan chain helps us to treat the output of the $ICOD$ of each flip-flop as a pseudo-output. This allows us to partition the circuit into multiple smaller instances based on the $ICOD$ of each flip-flop. For example, we can treat the $ICOD_{DFF1}$ as a standalone circuit consisting of six inputs ($I1 \rightarrow I6$) and one output ($m$). We can apply any input pattern to $I1 \rightarrow I6$ in functional mode and observe the value of $m$ by shifting out the content of $DFF1$ in scan mode. As $ICOD_{DFF1}$ contains only two key-gates, it would be easier to apply brute force attack on $ICOD_{DFF1}$ to extract these keys. The attacker can apply the same input pattern to the $ICOD_{DFF1}$ of an encrypted and an activated IC and observe $m$ to extract $K1$ and $K2$. Similarly, the keys, affecting the $ICOD$s of other flip-flops, can also be extracted by treating those $ICOD$s as separate independent logic circuits. 
	
	\begin{algorithm}[!ht]
		\SetKwInOut{Input}{Input}
		\SetKwInOut{Output}{Output}
		\SetKwInOut{Var}{Var}
		\scriptsize{
			\Input{Encrypted Netlist; Activated IC;}
			\Output{Key values;}
			\Begin{
				\For{All the scan flip-flops}{
					Find the input cone of dependency ($ICOD$)\;}
				\While{All the keys are not extracted}{
					\While{The $ICOD$ of all the scan flip-flops are not explored}{
						
						Find out the scan flip-flop $FF_{least}$ with the least number of key-gates in its $ICOD$\;
						\If{The $ICOD$ of $FF_{least}$ contains other flip-flops ($FF_j$)}{
							Operate the circuit in scan mode\;
							Treat the outputs of $FF_j$s as pseudo-inputs by uploading any values in $FF_j$s through scan chain\;
							Switch to functional mode\;
						}
						Apply input pattern in functional mode\;
						Switch to scan mode\;
						Shift out the content of the scan chain through scan out port\;
						Observe the output of the $ICOD$ of $FF_{least}$ of both encrypted and activated ICs\;
						Apply brute-force attack on $ICOD$ of $FF_{least}$ to extract the keys affecting it\;
					}   
					\If{Any key-gate does not belong to the $ICOD$ of any scan flip-flop}{
						Apply logic cone attack to extract the key\;
					}
				}
			}
		}
		\caption{\textit{Proposed Scan-based Attack}}\label{algo:scan_attack}
	\end{algorithm}
	\setlength{\textfloatsep}{0pt}
	Algorithm \ref{algo:scan_attack} outlines the proposed scan based attack on logic encryption. The attack starts with partitioning the circuit into several smaller sub-circuits based on the $ICOD$ of each flip-flop. Next, it searches the flip-flop ($FF_{least}$) with the least number of key-gates in its $ICOD$. The keys corresponding to that $ICOD$ are extracted using brute force attack. Typical logic cone based attack considers the $ICOD$ of any output to perform the attack. We can observe from Figure \ref{fig:our_attack} that the input cone of dependency $ICOD_{Oi}$ of the output $O_i$ contains more primary inputs, logic gates, and key-gates compared to the $ICOD$ of any flip-flop. As the complexity of brute force attack exponentially increases with the number of key-gates and the size of a circuit, it is much easier to apply the attack on the $ICOD$ of any flip-flop compared to that of any output. The attack iteratively searches for the flip-flop with the least number of key-gates ($FF_{least}$) in its $ICOD$ and extract those keys. A brute force attack on an $ICOD$ which contains only combinational elements is straightforward. However, if the $ICOD$ of $FF_{least}$ contains other flip-flops (say $FF_j$s), an attacker has to operate the circuit in scan mode to use the outputs of the $FF_j$s as pseudo-primary inputs before switching back to the functional mode of operation. For example, $DFF4$ contains $DFF2$ in its input cone of dependency $ICOD_{DFF4}$. Therefore, at the time of extracting the keys present in $ICOD_{DFF4}$, first, we operate the circuit in scan mode and upload any value to $DFF2$ and then switch to functional mode. This mutes the $ICOD_{DFF2}$ and reduces the $ICOD$ of $DFF4$ from $ICOD_{DFF4}$ to ($ICOD_{DFF4}$ - $ICOD_{DFF2}$). 
	
	The attack can be applied to any state-of-the-art logic encryption technique provided the circuit contains flip-flops with DfT infrastructure for infield testing. The complexity of the attack does not depend on the total number of key-gates and primary inputs. Rather it depends on the number of key-gates present in the largest $ICOD$ of any flip-flop and the number of primary inputs affecting that $ICOD$. If the largest $ICOD$ of any flip-flop contains $K1$ key-gates and $M1$ primary inputs, the complexity of the attack reduces from $O(2^{M+K})$ to $O(2^{M1+K1})$. In general, $K1 \ll K$ and $M1 < M$, therefore, it becomes easy to apply the attack on a large complex circuit, encrypted with a sufficiently large number of keys.
	
	To examine the effectiveness of scan based attack, we have applied the attack on several ISCAS'89 and ITC'99 benchmarks, which are encrypted using typical XOR/XNOR based encryption strategy (128-bit key). Table \ref{tab:scan_attack_complexity} reports the complexities of both scan-based and brute force attacks (in the format, scan attack complexity / brute force complexity) on different ISCAS'89 and ITC'99 benchmarks. We observe that the scan based attack can drastically reduce the attack complexity. For example, the brute force attack complexity on $s15850$ is $2^{128+77}$ = $2^{205}$, which can be reduced down to $2^{62}$ by exploiting the scan chains of the design. This makes several other attacks feasible, which would not have been possible otherwise. This attack shows the importance of introduction of encryption in scan chains, which has not been considered in the literature.

	\begin{table}[!ht]
		\centering
		\caption{Complexity of scan attack on several ISCAS'89 and ITC'99 benchmarks (K = 123 for s9234, K = 128 for others)}
		\label{tab:scan_attack_complexity}
		\begin{tabular}{|c|c|c|c|}
			\hline
			
			\textbf{\begin{tabular}[c]{@{}c@{}}Circuit \\ Name\end{tabular}} & \textbf{\begin{tabular}[c]{@{}c@{}}Attack complexity \\ Scan / Brute Force\end{tabular}} & \textbf{\begin{tabular}[c]{@{}c@{}}Circuit \\ Name\end{tabular}} & \textbf{\begin{tabular}[c]{@{}c@{}}Attack complexity \\ Scan / Brute Force\end{tabular}} \\ \hline
			\textbf{$s5378$} & $2^{66}$ / $2^{163}$ & \textbf{$b17$} & $2^{61}$ / $2^{165}$ \\ \hline
			\textbf{$s9234$} & $2^{51}$ / $2^{159}$ & \textbf{$b18$} & $2^{47}$ / $2^{165}$\\ \hline
			\textbf{$s13207$} & $2^{56}$ / $2^{190}$ & \textbf{$b19$} & $2^{51}$ / $2^{152}$\\ \hline
			\textbf{$s15850$} & $2^{62}$ / $2^{205}$ & \textbf{$b20$} & $2^{64}$ / $2^{160}$\\ \hline
			\textbf{$s38417$} & $2^{69}$ / $2^{156}$ & \textbf{$b21$} & $2^{58}$ / $2^{160}$\\ \hline
			\textbf{$s38584$} & $2^{74}$ / $2^{166}$ & \textbf{$b22$} & $2^{53}$ / $2^{160}$\\ \hline
		\end{tabular}
	\end{table}
	
	\vspace{-0.3cm}
	\section{Encrypt Flip-flop: A New Logic Encryption Strategy}
\begin{figure}[!ht]
		\centering
		\includegraphics[scale = 0.32]{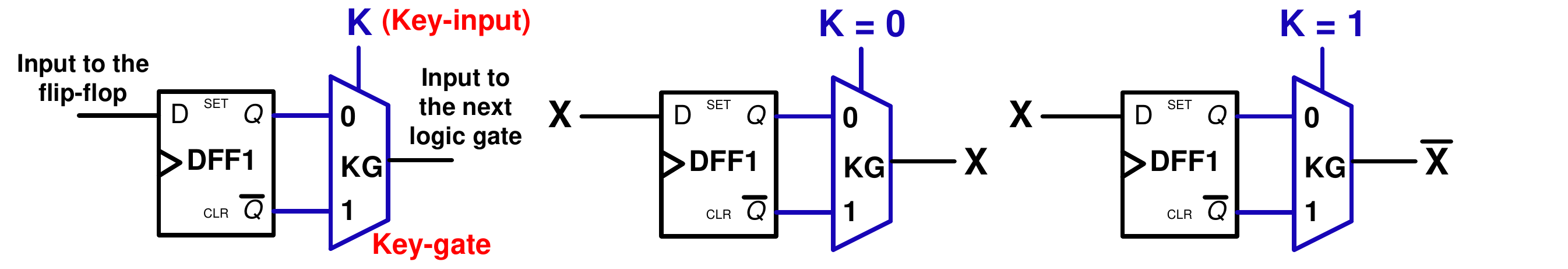}
		\caption{The basic idea of \textit{Encrypt Flip-Flop} strategy}
		\label{fig:our_logic_encryption}
		\vspace{-0.5cm}
	\end{figure}

	In the previous section, we have shown that any state-of-the-art logic encryption technique is vulnerable to scan based attack. To restrict the leakage of information through the scan chain, we propose to introduce obfuscation in the scan chain itself. In this section, we propose a new logic encryption strategy called \textit{Encrypt Flip-Flop}, which encrypts the outputs of the flip-flops of a sequential design. Flip-flops produce two outputs, $Q$, and $\overline{Q}$. Generally, one of these two outputs (either $Q$ or $\overline{Q}$) is connected to the next logic level, while the other remains unconnected. We propose to encrypt the outputs of some of the flip-flops by inserting MUXes in front of them. The two inputs of the MUX are connected to the $Q$ and $\overline{Q}$ lines of the flip-flop, and its output is connected to the next logic level. The MUX acts as a key-gate, while the select line of the MUX acts as the key. Either $Q$ or $\overline{Q}$ line passes to the next logic gate depending upon the value of the select line. Figure \ref{fig:our_logic_encryption} depicts the basic idea of our proposed \textit{Encrypt Flip-Flop} strategy. If the input to the flip-flop is $X$, a key-input of $K = 0$ passes $X$ to the output of the key-gate, while a value of $K = 1$ passes $\overline{X}$. 
	
\begin{figure}[!ht]
		\centering
		\includegraphics[scale = 0.24]{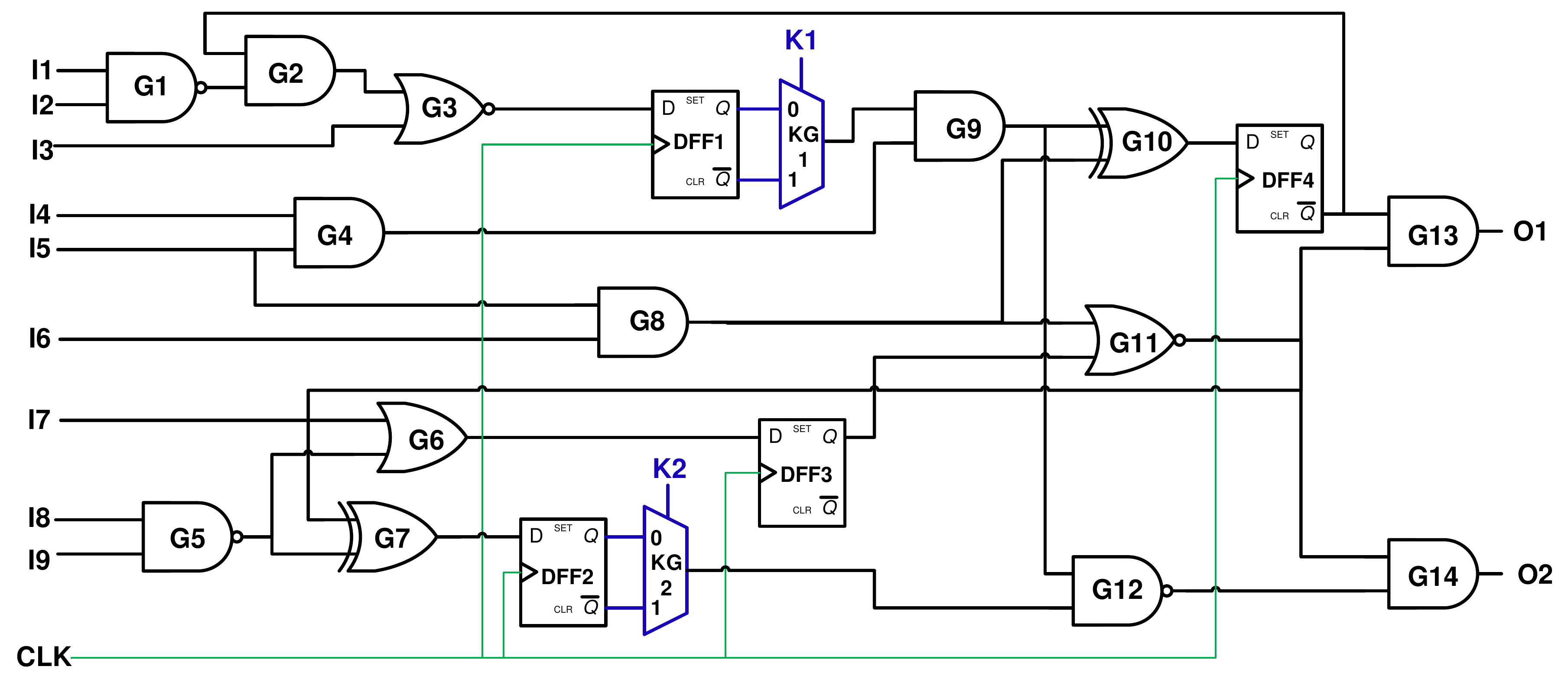}
		\caption{Encrypting the example circuit of Figure \ref{fig:example}(a) using \textit{Encrypt Flip-Flop} strategy}
		\label{fig:new_example}
		\vspace{-1cm}
	\end{figure}

	A designer can selectively opt to pass a $\overline{Q}$ / $Q$ value to the output of a key-gate, where an unencrypted design should propagate a $Q$ / $\overline{Q}$ value to the next logic level. The inversion introduced by this process can be bubble pushed further deep into the circuit using de Morgan's law. Therefore, any key-gate can have a key value of either '0' or '1'. A wrong key-input propagates an inverted input to the next logic level, which leads to an erroneous functionality of the circuit. It may be noted that MUX-based encryption has also been proposed in \cite{rajendran2015fault}. However, in \cite{rajendran2015fault}, the two inputs (true and false lines) of a key-gate (MUX) may carry same value (either 0 or 1) for certain input combinations. In such scenarios, even a wrong key propagates correct value to the next logic level. This situation never occurs in our case as we consider $Q$ and $\overline{Q}$  of a flip-flop as the two inputs of a key-gate. Unlike \cite{zhang2016practical}, our method does not use an extra inverter for each key-gate to ensure two different values in the two inputs of a key-gate. Figure \ref{fig:new_example} shows an example encryption of the circuit of Figure \ref{fig:example}(a) using our proposed \textit{Encrypt Flip-Flop} strategy. In general, a sequential circuit of medium size contains hundreds of flip-flops, while the number of flip-flops in a larger design can be in the order of thousands. Depending upon the permissible area constraint, a designer can encrypt a sufficiently large number of flip-flops of a design. However, random selection of flip-flops for encryption may expose a design to logic cone based attacks. Proper selection of flip-flops for encryption plays an important role to ensure the quality of security offered by the encrypted design.

	\subsection{Selection Of Flip-flops For Encryption}
	We have observed in Section \ref{sec:evo} that the vulnerability of a circuit against logic cone based attack increases if some outputs of the circuit do not contain a sufficient number of key-gates in their input cone of dependency. Therefore, the primary focus of our key-gate location selection process is to confine the effects of all the key-gates to a limited number of outputs, and at the same time, ensuring that each of the affected outputs contains all the key-gates in its input cone of dependency. Let us assume, a circuit consists of $M$ inputs, $O$ outputs and $L$ flip-flops. We would like to encrypt the circuit using a $K$-bit key. We will select $K$ out of $L$ flip-flops, which satisfy the following criterion. 
	
	\vspace{0.3cm}
	\fbox{\begin{minipage}{23em}
			\textbf{Each of the outputs, affected by any of these $K$ flip-flops, must contain all of these $K$ flip-flops in its input cone of dependency.}
		\end{minipage}}
		\vspace{0.3cm}
		
		We encrypt these $K$ flip-flops by inserting a MUX in front of each of them. The select lines of these MUXes act as key-inputs. The process ensures that the input cone of dependency of any output contains either all or none of the key-gates. None of the outputs contains only few key-gates in its input cone of dependency. Thus, an attacker does not get an opportunity to minimize the effort of brute force search by employing logic cone based attack.

		Algorithm \ref{algo:key-location} describes the proposed strategy of selecting the flip-flops for encryption. The process starts by finding the flip-flops present in the input cone of dependency ($ICOD$) of all the $O$ outputs of a circuit. Next, we select $O'$ outputs ($O' \leq O$) with the largest number of flip-flops in their overlapping input cone of dependency ($ICOD_{overlap}$). Let, $ICOD_{overlap}$ contains $L1$ ($L1 \leq L$) flip-flops. We name this set of $L1$ flip-flops as $L_{overlap}$. Each of these $L1$ flip-flops affects all the $O'$ outputs. However, some of these $L1$ flip-flops may also affect some outputs from the set of $O-O'$ outputs. Let, $L2$ flip-flops (out of the $L1$ flip-flops) affect any of the $O-O'$ outputs. As none of these $O-O'$ outputs contains a sufficient number of key-gates in its input cone of dependency, the keys corresponding to these $L2$ flip-flops can be extracted by applying logic cone based attack on these $O-O'$ outputs. Therefore, these $L2$ flip-flops are not suitable candidates for encryption. We name this set of $L2$ flip-flops as $L_{weak}$. We exclude these $L2$ flip-flops from $L_{overlap}$ and create a new set ($L_{strong}$) of $L3  = L1 - L2$ flip-flops. None of the elements of the set $L_{strong}$ affects any output other than the $O'$ outputs. All of these $L3$ flip-flops are the potential candidates for encryption. We can select any $K$ flip-flops from these $L3$ flip-flops and encrypt them by inserting MUX in front of them. Encryption of these $K$ flip-flops ensures that each of the $O'$ outputs includes all the $K$ key-gates, and other $O-O'$ outputs contain no key-gate in their input cone of dependency. 
		
		\begin{algorithm}[!ht]
			\SetKwInOut{Input}{Input}
			\SetKwInOut{Output}{Output}
			\SetKwInOut{Var}{Var}
			\scriptsize{
				\Input{Original Netlist; Key size ($K$);}
				\Output{Encrypted Netlist;}
				\Begin{
					\For{Each of the O outputs}{
						Find the flip-flops present in its input cone of dependency ($ICOD$)\;}
					Form a set of $O'$ outputs with the largest number of flip-flops ($L_{overlap}$) in the overlapping input cone of dependency ($ICOD_{overlap}$)\;
					\For{each element of $L_{overlap}$}{
						\If{The flip-flop affects any output other than the O' outputs}{
							Include the flip-flop in the set $L_{weak}$\;
						}
					}
					Form a new set $L_{strong}$ = $L_{overlap}$ - $L_{weak}$\;
					Select $K$ flip-flops from the set $L_{strong}$\;
					Insert a MUX in front of these $K$ flip-flops\;
				}
			}
			\caption{\textit{Key-gate Location Selection}}\label{algo:key-location}
		\end{algorithm}
		\setlength{\textfloatsep}{0pt}
		
		\begin{figure}[!ht]
			\centering
			\includegraphics[scale = 0.18]{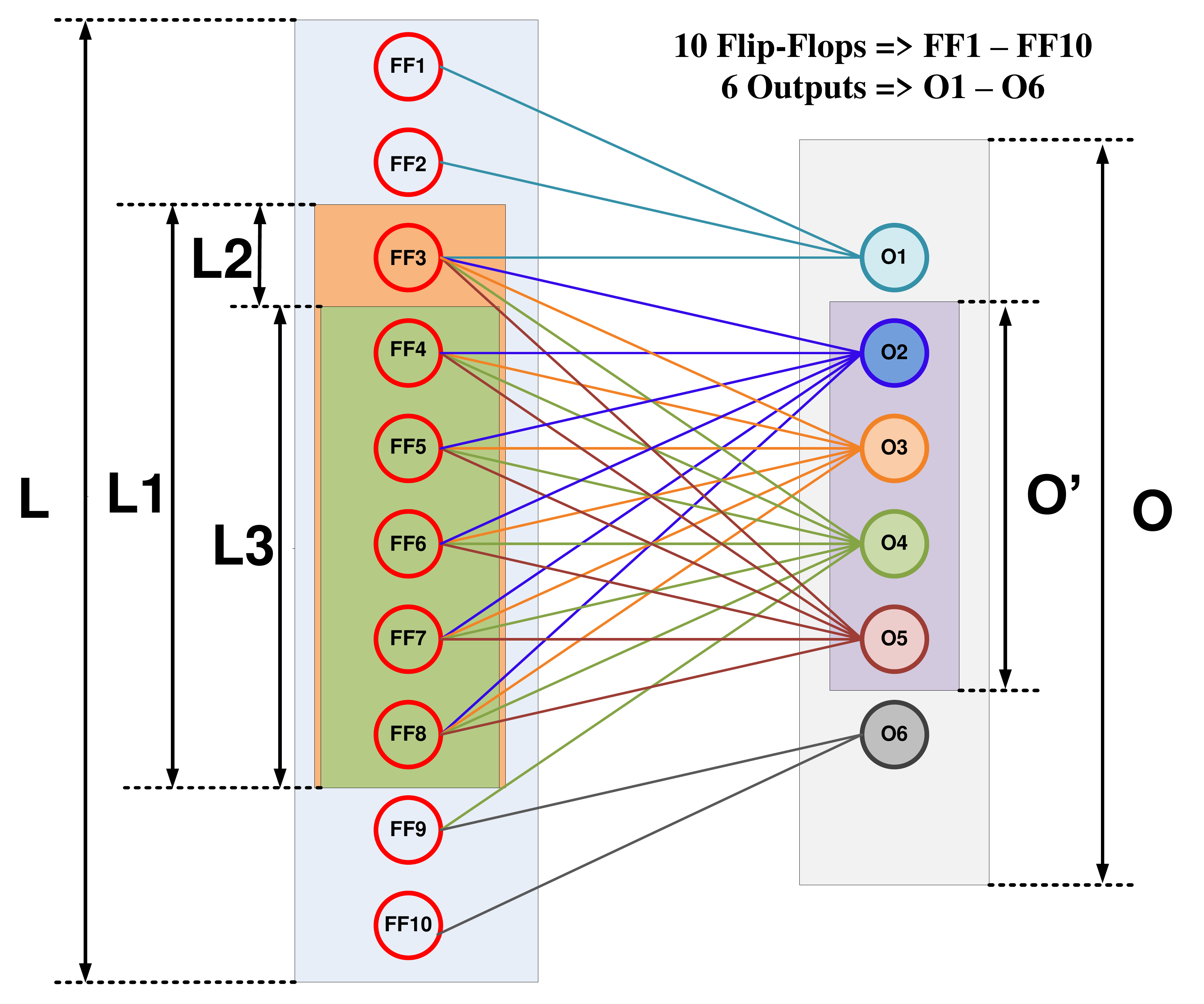}
			\caption{Selection of flip-flops for encryption based on Algorithm \ref{algo:key-location}}
			\label{fig:overlap}
			\vspace{-0.5cm}
		\end{figure}
		
		The process can be explained using the example of Figure ~\ref{fig:overlap}. Let, the example circuit contains 10 flip-flops, namely $FF1$ to $FF10$ and 6 outputs, namely $O1$ to $O6$. Input cone of dependency of all the outputs is represented using a graph (Figure ~\ref{fig:overlap}). Each of the flip-flops and the outputs is represented by a node in the graph, while an edge between an output and a flip-flop signifies that the flip-flop belongs to the input cone of dependency of that output. For example, the input cone of dependency of output $O1$ contains the flip-flops $FF1$, $FF2$, and $FF3$. It may be observed from the figure that the outputs $O2$, $O3$, $O4$, and $O5$ have the flip-flops $FF3$, $FF4$, $FF5$, $FF6$, $FF7$, and $FF8$ common in their $ICOD$s and these $ICOD$s construct the largest overlapping $ICOD$. Therefore, according to the example, $ICOD_{overlap}$ contains the outputs $O2$, $O3$, $O4$, and $O5$ and $L_{overlap}$ contains the flip-flops $FF3$, $FF4$, $FF5$, $FF6$, $FF7$, and $FF8$. Please note that the flip-flop $FF3$ also affects the output $O1$ which is not an element of the list $O'$. If we encrypt the output of $FF3$, it would be easier to extract the corresponding key by applying logic cone based attack on the output $O1$. Therefore, the flip-flop $FF3$ belongs to the set $L_{weak}$. We exclude $FF3$ from $L_{overlap}$ and construct the set $L_{strong}$ which contains the flip-flops $FF4$, $FF5$, $FF6$, $FF7$, and $FF8$. It may be noted that every element of $L_{strong}$ affects only the outputs of the list $O'$. Encryption of the flip-flops of the set $L_{strong}$ ensures protection against logic cone based attack.

		\section{Security Analysis}
		In this section, we evaluate the security of our proposed approach against several attacks proposed in the literature. We have already discussed about sustainability against logic cone analysis based attack \cite{lee2015improving}. Therefore, we mainly focus on other security threats like path sensitization attack \cite{yasin2015improving}, scan based attack, SAT attack\cite{subramanyan2015evaluating} etc.
		
		\subsection{Security Evaluation Against Path Sensitization Attack}
		In this section, we examine whether a path sensitization attack can be performed on a design encrypted using our proposed strategy. Path sensitization attack on the proposed logic encryption is slightly different from the attack proposed in \cite{yasin2015improving}.  Unlike \cite{yasin2015improving}, we cannot directly propagate a key-value to an output. Rather, we have to apply a specific value to the input of an encrypted flip-flop and propagate that value to the output by selecting a key-value (either 0 or 1). Comparison of the output with the output of an activated IC decides whether the applied key is correct or wrong. Therefore, to perform path sensitization attack, it is important to control the input of an encrypted flip-flop. Input of a flip-flop can be controlled in two ways. Input of few flip-flops can be directly controlled by manipulating the primary inputs (although the number of such flip-flop is very less). Inputs of the rest of the flip-flops can also be controlled, provided an activated IC has the scan facility for the purpose of in-field testing. This is very common as most of the ICs have the DfT (Design-for-Testability) infrastructure. 
		
\begin{figure}[!ht]
			\centering
			\includegraphics[scale = 0.23]{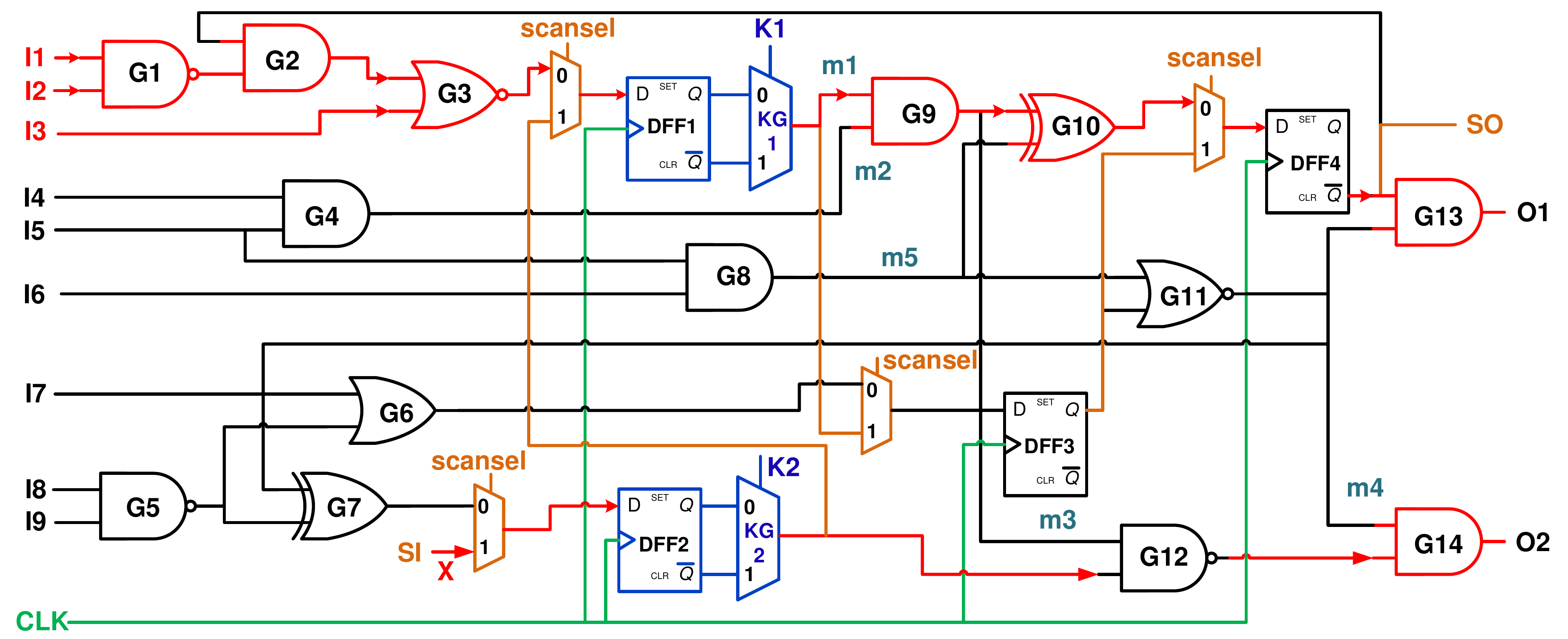}
			\caption{Path sensitization attack on \textit{Encrypt Flip-Flip}, Case Study I}
			\label{fig:path_sensitization_attack1}
			\vspace{-0.5cm}
		\end{figure}

		\textbf{Case I:} To demonstrate the attack, we consider the same example circuit of Figure ~\ref{fig:new_example} which is encrypted using our proposed method. We assume that the circuit has the DfT infrastructure. Therefore, we convert the flip-flops into scan flip-flops (Figure ~\ref{fig:path_sensitization_attack1}) and connect them in a chain. Our objective is to extract the keys $K1$ and $K2$. It may be noted that the input of $DFF1$ can be directly controlled from the primary inputs ($I1$, $I2$, and $I3$). The value of $DFF1$ can be propagated to the output $O1$ by selecting any value of $K1$ in the encrypted netlist and setting the lines $m2$ ($m2$ = 1), $m5$ ($m5$ = 0), and $m4$ ($m4$ = 1) to their non-controlling values by manipulating the primary inputs. We compare the value of $O1$ of the activated IC for the same input vector. If both the outputs are same, we decide that the selected value of $K1$ is correct, else the correct value of $K1$ should be the other one. On the other hand, input of $DFF2$ cannot be directly controlled from the primary inputs. Therefore, we switch from functional mode to scan mode and insert a value 'X' (either '0' or '1') to the input of $DFF2$. Next, we switch from scan mode to functional mode and propagate the value of 'X' to the output $O2$ and extract $K2$ in the similar fashion as $K1$. 
		
\begin{figure}[!ht]
			\centering
			\includegraphics[scale = 0.21]{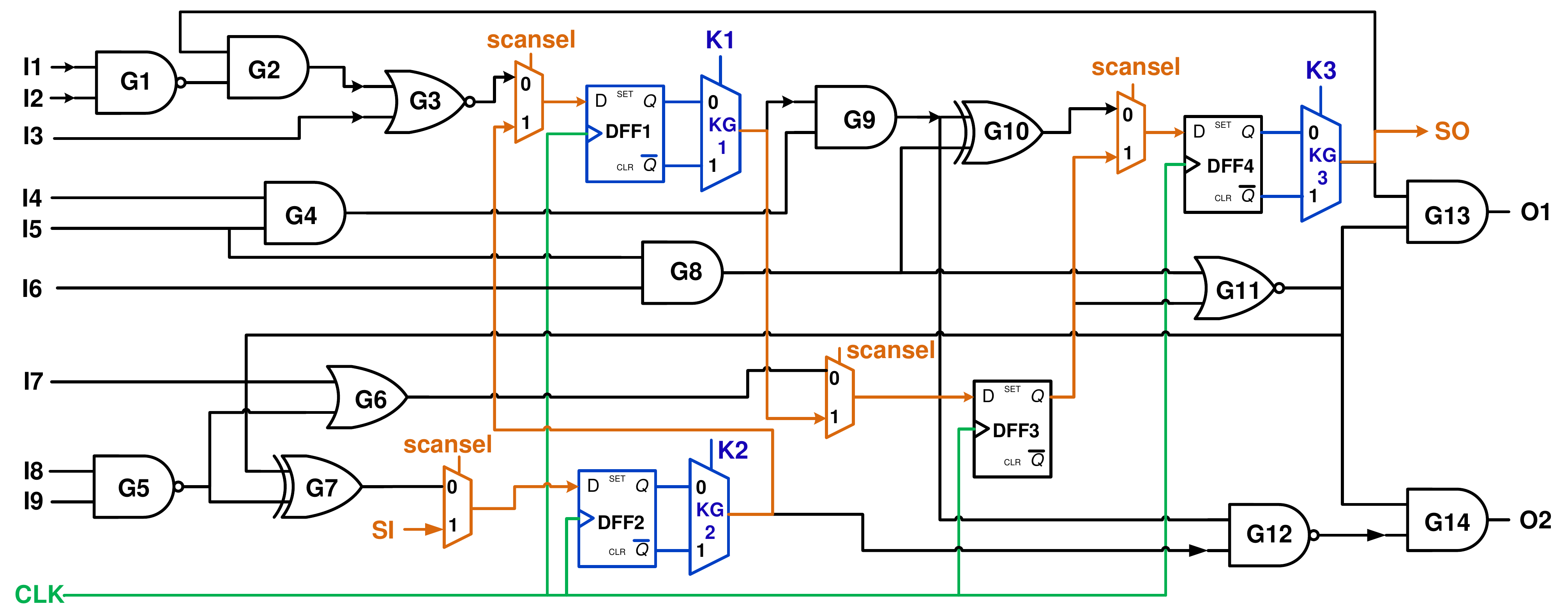}
			\caption{Path sensitization attack on \textit{Encrypt Flip-Flip}, Case Study II}
			\label{fig:path_sensitization_attack2}
			\vspace{-0.5cm}
		\end{figure}

		\textbf{Case II:} Now, let us consider the scenario of Figure ~\ref{fig:path_sensitization_attack2}. Here, we encrypt the flip-flop $DFF4$ as well. The input of $DFF4$ cannot be controlled from the primary inputs. Hence, we try to control the input of $DFF4$ via the scan chain. However, it may be noted that the scan input ($SI$) of flip-flop $DFF2$ is only externally accessible in an activated IC. All the scan inputs are applied only through this $SI$ line. In normal scenario, where none of the flip-flops are encrypted, we can easily upload our desired value to the input of $DFF4$ by applying that value to $SI$ and scan shifting the value to the input of $DFF4$. However, in our encrypted design, if we do not know the values of the keys $K1$ and $K2$, we cannot figure out which value is actually propagating to the next scan flip-flop of an encrypted flip-flop. Therefore, it is not possible to control the input of $DFF4$, hence the key $K3$ cannot be extracted using path sensitization attack. 
		
		\begin{figure*}[!ht]
			\centering
			\includegraphics[scale = 0.248]{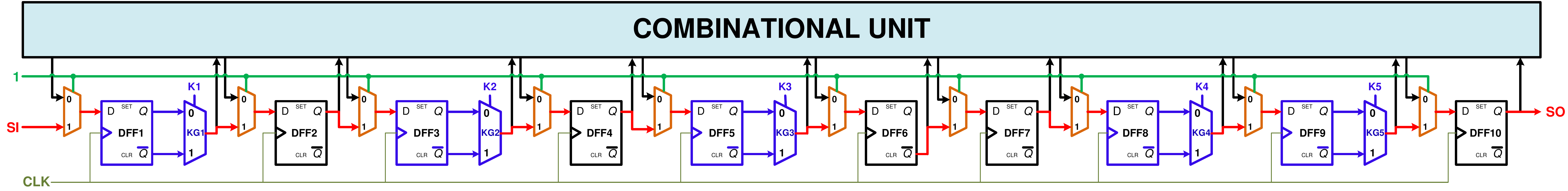}
			\caption{An example scan chain, encrypted using \textit{Encrypt Flip-Flop} technique}
			\label{fig:scan_chain_example}
			\vspace{-0.6cm}
		\end{figure*}
		
		\textbf{Observation:} From the above example, one can observe that our proposed encryption strategy is vulnerable to path sensitization attack only if the input of an encrypted flip-flop is either directly controllable from the primary inputs or connected to the scan in ($SI$) line of the scan chain (i.e. the first flip-flop of a scan chain). In general, very few flip-flops can be controlled from the primary inputs. On the other hand, the input of only one encrypted flip-flop of a scan chain can be controlled externally. Sensitization of that key can also be restricted if sufficiently large number of key-gates interfere in the sensitization path of that key. This can be done by selecting an encrypted flip-flop, which has high interference in its sensitization path, as the first flip-flop of a scan chain. Therefore, path sensitization attack cannot be performed on our proposed encryption strategy, if the flip-flops, whose inputs are not controllable from the primary inputs, are selected for the purpose of encryption.  
		
		\subsection{Security Evaluation Against Scan-based Attack}
		Earlier, we have shown that scan paths can be exploited to reduce the efforts of different attacks on logic encryption. Unlike other encryption techniques, our proposed strategy does not include any key-gate in the input cone of dependency of a scan flip-flop, thus, prevents an attacker from using the scan chains to apply divide and conquer based scan attack, proposed in Section \ref{leb:scan_attack}. However, as we encrypt the scan flip-flops themselves, scan shifting can be a potential source of leakage of key information. In this section, we investigate the resilience of our proposed \textit{Encrypt Flip-Flop} technique against scan based attack.  For this purpose, we consider an example circuit with a scan chain containing ten flip-flops (Figure ~\ref{fig:scan_chain_example}), five ($DFF1$, $DFF3$, $DFF5$, $DFF8$, and $DFF9$) of them are encrypted using \textit{Encrypt Flip-Flop} technique. We consider two scenarios where we apply the same input scan vector under two different key setups and analyze the data extracted by shifting out the contents of the scan chain. 
		
\begin{figure}[!ht]
			\centering
			\includegraphics[scale = 0.45]{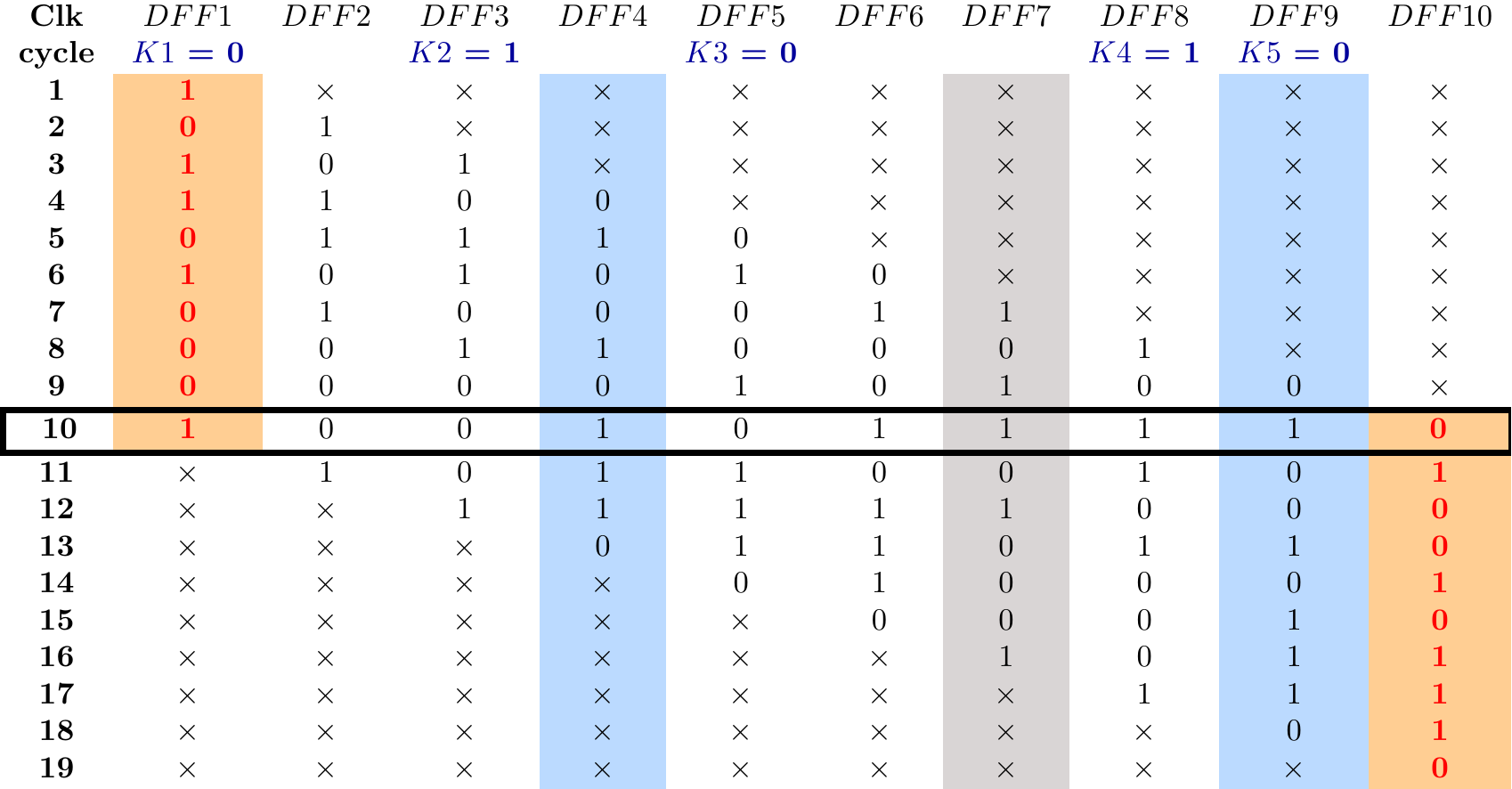}
			\caption{Shifting of scan data through the example scan chain of Figure \ref{fig:scan_chain_example}, considering the key "01010"}
			\label{fig:scan_attack1}
			\vspace{-0.7cm}
		\end{figure}

		\textbf{Case I:} 
		Let us assume the correct values of five keys of the example Figure ~\ref{fig:scan_chain_example} be $K1 = 0$, $K2 = 1$, $K3 = 0$, $K4 = 1$, and $K5 = 0$. We upload the scan vector "1011010001" into the scan chain through the scan in ($SI$) port. Figure ~\ref{fig:scan_attack1} shows how this scan data shifts through the scan chain for the chosen key-values. As the values of the keys $K2$ and $K4$ are '1', inverted values get shifted to the next flip-flops (i.e. $DFF4$ and $ DFF9$). Another inversion is caused by the connection between $\overline{Q}$ output of $DFF6$ to the input of $DFF7$. Thus a scan vector faces three inversions during the scan shift. However, an attacker does not have direct access to the content of the intermediate flip-flops ($DFF2 \rightarrow DFF9$). Shifting out the content of the intermediate scan flip-flops through the scan out port (i.e. the output of $DFF10$) is the only option to observe these values.  Figure ~\ref{fig:scan_attack1} shows that under this particular key setup, the scanned out vector is exactly the inverted one of the uploaded input vector. 
		
\begin{figure}[!ht]
			\centering
			\includegraphics[scale = 0.45]{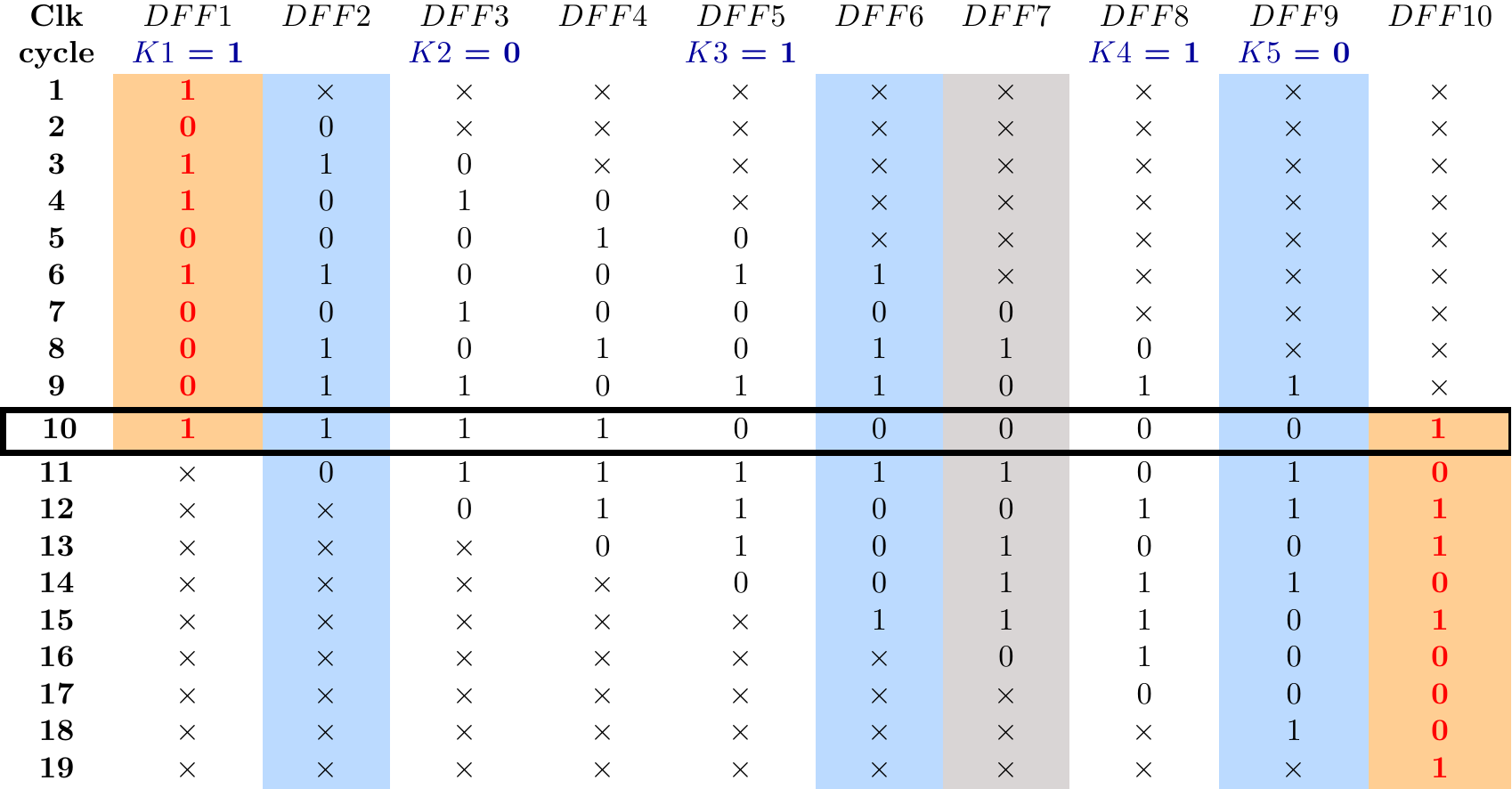}
			\caption{Shifting of scan data through the example scan chain of Figure \ref{fig:scan_chain_example}, considering the key "10110"}
			\label{fig:scan_attack2}
			\vspace{-0.3cm}
		\end{figure}

		\textbf{Case II:}
		Now, let us consider a different key ($K1 = 1$, $K2 = 0$, $K3 = 1$, $K4 = 1$, and $K5 = 0$) and the same input scan vector "1011010001". Figure ~\ref{fig:scan_attack2} shows the scan shift under this particular key setup. In this case, we can observe that the scanned out vector is exactly the same as the uploaded input vector.
		
		\textbf{Observation:}
		We have observed that in the first scenario, output scan vector gets inverted, while in the second one, it remains unaltered. Please note that a scan vector gets inverted in the process of shifting through a scan flip-flop if either the $\overline{Q}$ output of an unencrypted flip-flop is connected to the next flip-flop or a '1' value of the key-input of an encrypted flip-flop is selected. The total number of these two cases decides the number of times a scan vector gets inverted during the shifting through a scan chain. The output vector remains unaltered for an even number of inversions, while an odd number of inversions inverts the output. For example, the $\overline{Q}$ output of only $DFF6$ is connected to the next flip-flop in the scan chain structure of Figure ~\ref{fig:scan_chain_example}, thus a single inversion is caused by all the unencrypted flip-flops. In scenario I, two key-bits are '1', which indicates the scan vector gets inverted twice by all the encrypted flip-flops. Thus, the total number of inversions is three and the final output gets inverted. Under this particular example scan infrastructure, an inverted output can also be observed in the cases where an even number of key-bits are '1'. Similarly, in scenario II, three key-bits are '1', which implies that the scan vector gets inverted for a total of four times and finally produces non-inverted output. Any key with an odd number of '1's produces the same output. This is true for any input scan vector. Please note that the number of connections from the $\overline{Q}$ output of an unencrypted flip-flop to the next flip-flop can be easily identified by observing the scan chain structure. However, it is not possible to figure out the number of '1's present in a key by observing the output vector. Therefore, an output scan vector can only identify whether a key contains an even or an odd number of '1's in it. It does not reveal any information regarding the number of '1's present in a key. A $K$ bit key has $2^{K-1}$ possible combination of keys, where an even number of bits are '1', and $2^{K-1}$ possible combination of keys, where an odd number of bits are '1'. For example, out of the 32 possible key combinations, 16 key combinations of the Figure ~\ref{fig:scan_chain_example} have either an even or an odd number of '1's. Therefore, by observing the type of scan output, an attacker can eliminate $2^{K-1}$ possible combination of keys which can only reduce the complexity of a brute force attack from O($2^{M+K}$) to O($2^{M+(K-1)}$).

\begin{figure}[!ht]
			\centering
			\includegraphics[scale = 0.134]{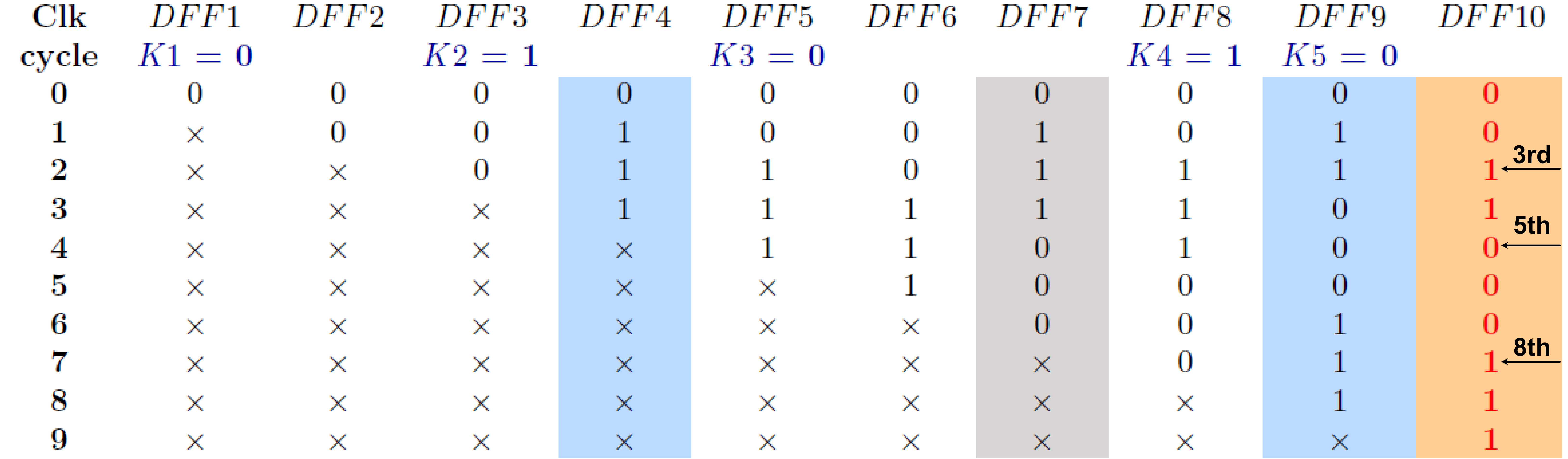}
			\caption{An example of reset-and-scan attack on the example scan chain of Figure \ref{fig:scan_chain_example}, considering the key "01010"}
			\label{fig:reset_and_scan}
			\vspace{-0.6cm}
		\end{figure}

		\textbf{Reset-and-Scan Attack on Encrypt Flip-Flop:}
		Simple scan operation does not reveal any key, as the logic values of the scan cells before scan operation remain unknown to the attacker. However, a design with global reset is vulnerable to reset-and-scan attack. Figure \ref{fig:reset_and_scan} demonstrates the attack on the example scan chain of Figure \ref{fig:scan_chain_example}, considering the key to be "01010". A global reset configures all the flip-flops to a known logic value (i.e. logic value '0'). A subsequent scan operation inverts the contents of the scan cells depending upon the key values, and finally, the scanned out vector reflects these inversions. By observing the inversion positions in the scanned out vector, the attacker can identify the locations of the keys which are responsible for those inversions. For example, the scan out vector in Figure \ref{fig:reset_and_scan} experiences three inversions at $3^{rd}$, $5^{th}$ and $8^{th}$ positions. These three inversions are caused by the key $K4 = 1$, the $\overline{Q}$ line of $DFF6$ and the key $K2 = 1$, respectively. It may be noted that the respective positions of $DFF8$ (associated with key-gate $KG4$), $DFF6$ and $DFF3$ (associated with key-gate $KG2$) are $3^{rd}$, $5^{th}$ and $8^{th}$ from the scan out port, which are same as the inversion positions.
		The inversion in the $5^{th}$ position is created by the $\overline{Q}$ line of $DFF6$, which can be identified from the netlist of the design. The other two inversions in $3^{rd}$ and $8^{th}$ positions reveal the values of $K4$ and $K2$ as logic '1'. 
		No other inversion in the scanned out vector suggests that all other key-bits are '0'. This way, a scan operation, immediately after a global reset can reveal all the keys.
		
\begin{figure}[!ht]
			\centering
			\includegraphics[scale = 0.16]{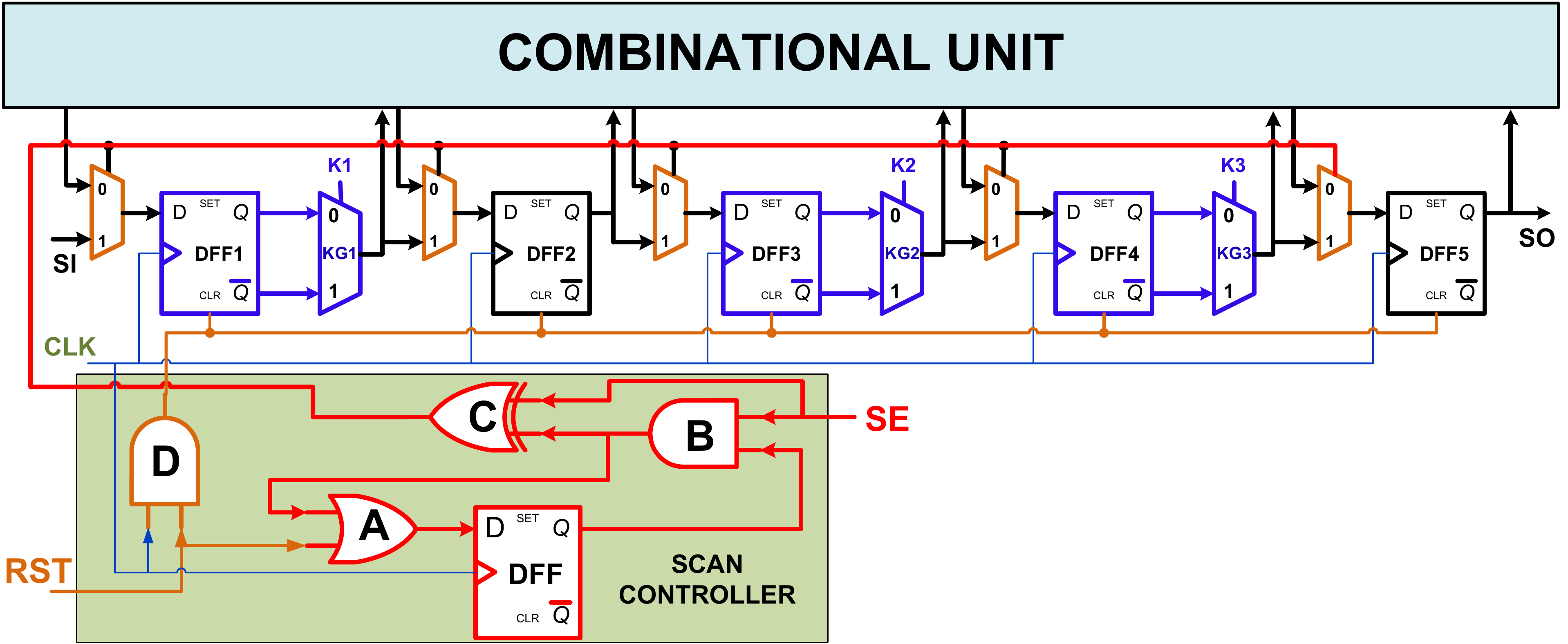}
			\caption{Scan controller to prevent reset-and-scan attack}
			\label{fig:reset_and_scan_prevention}
			\vspace{-1cm}
		\end{figure}

			\textbf{Countermeasure Against Reset-and-Scan Attack:}
			To prevent the reset-and-scan attack, a designer must restrict a scan operation, immediately after a global reset. To do so, we propose to introduce a scan controller into the design. Figure \ref{fig:reset_and_scan_prevention} shows the architecture of the proposed scan controller. Instead of applying the scan enable (SE) input directly to the select lines of the scan MUXs, we apply the SE and reset (RST) inputs to the scan controller. The output of the scan controller is applied to the select lines of the scan MUXs. The scan controller operates as follows. A global reset makes RST = 1.  In the first clock cycle, the output of the AND gate D resets all the flip-flops. In the next clock cycle, the flip-flop $DFF$ propagates the value of RST to the input of the AND gate $B$.  For an immediate scan operation, the value of SE needs to be '1'.  This sets the output of $B$ to the logic value '1'. Although RST becomes '0' from the next cycle onward, as the output of $B$ is fed back to the $DFF$ via the OR gate $A$, the output of $B$ remains '1'. As the values of both $B$ and SE are '1', the output of the XOR gate $C$ becomes '0'. This disables the scan operation. The value of $B$ remain '1' until SE becomes '0'. Therefore, the scan controller does not allow the attacker to perform a scan operation, immediately after a global reset. However, when we switch from normal mode to scan mode, only the value of SE becomes '1', and the value of RST remains '0'. Therefore, the value of $B$ remains '0' and the output of $C$ becomes '1', which enables the scan operation. As the integration of the scan controller restricts scan after a reset operation, an attacker cannot interpret the keys by observing the scanned out response, which helps us to prevent the reset-and-scan attack. 
		
		\begin{table*}[!ht]
			\centering
			\caption{Details of different benchmark circuits and execution times of the proposed \textit{Encrypt Flip-flip} strategy}
			\label{tab:circuit_details}
			\begin{tabular}{|c|c|c|c|c|c|c|c|c|}
				\hline
				\textbf{\begin{tabular}[c]{@{}c@{}}Circuit \\ Name\end{tabular}} & \textbf{\begin{tabular}[c]{@{}c@{}}\# \\ Inputs\end{tabular}} & \textbf{\begin{tabular}[c]{@{}c@{}}\#\\ Outputs\end{tabular}} & \textbf{\begin{tabular}[c]{@{}c@{}}\#\\ Gates\end{tabular}} & \textbf{\begin{tabular}[c]{@{}c@{}}\#\\ DFFs\end{tabular}} & \textbf{\begin{tabular}[c]{@{}c@{}}\#Candidate \\ DFFs\end{tabular}} & \textbf{\begin{tabular}[c]{@{}c@{}}\#Affected \\ Outputs\end{tabular}} & \textbf{\begin{tabular}[c]{@{}c@{}}Affected Output\\ Coverage (\%)\end{tabular}} & \textbf{\begin{tabular}[c]{@{}c@{}}Encryption \\ Time\end{tabular}} \\ \hline
				\textbf{s5378} & 35 & 49 & 2958 & 179 & 166 & 49 & 100 & 2.75 sec \\ \hline
				\textbf{s9234} & 36 & 39 & 5808 & 211 & 123 & 19 & 48.7 & 2.98 sec \\ \hline
				\textbf{s13207} & 62 & 152 & 8589 & 638 & 377 & 72 & 47.3 & 52.91 sec \\ \hline
				\textbf{s15850} & 77 & 150 & 10306 & 534 & 447 & 27 & 18 & 45.2 sec \\ \hline
				\textbf{s38584} & 38 & 304 & 20679 & 1426 & 1425 & 268 & 88.15 & 43 min 7 sec \\ \hline
				\textbf{s38417} & 28 & 106 & 23815 & 1636 & 1448 & 33 & 31.13 & 8 min 5 sec \\ \hline
				\textbf{b17} & 37 & 97 & 29267 & 1415 & 1321 & 30 & 30.92 & 39 min 8 sec \\ \hline
				\textbf{b18} & 37 & 23 & 97569 & 3320 & 3191 & 20 & 86.95 & 113 min 46 sec \\ \hline
				\textbf{b19} & 24 & 30 & 196855 & 6642 & 6157 & 27 & 90 & 695 min 9 sec \\ \hline
				\textbf{b20} & 32 & 22 & 17648 & 490 & 449 & 22 & 100 & 4 min 4 sec \\ \hline
				\textbf{b21} & 32 & 22 & 17972 & 490 & 449 & 22 & 100 & 4 min 6 sec \\ \hline
				\textbf{b22} & 32 & 22 & 26195 & 735 & 641 & 22 & 100 & 4 min 30 sec \\ \hline
			\end{tabular}
			\vspace{-0.5cm}
		\end{table*}

		\subsection{Security Evaluation Against SAT-based attack}
		SAT-based attack is applicable only on combinational circuits. However, the attack can also be performed on sequential circuits in presence of DfT architecture \cite{subramanyan2015evaluating}. To perform the attack, an attacker requires full controllability and observability of all the scan cells, which is very much possible in traditional scan chains. However, this is not possible in our proposed \textit{Encrypt Flip-Flop} technique as we restrict the controllability and observability of the internal scan cells. To illustrate our claim, we consider the same example encrypted scan configuration of Figure \ref{fig:scan_chain_example}. We apply the input scan vector "1011010001" under the key configuration of $K1 = 0$, $K2 = 1$, $K3 = 0$, $K4 = 1$, and $K5 = 0$. It takes ten clock cycles to upload the scan vector to all the scan cells. It can be observed from the Figure \ref{fig:scan_attack1} that, after ten clock cycles, the contents of the scan cells are "1001011110". If we apply the same input scan vector under the key configuration of $K1 = 1$, $K2 = 0$, $K3 = 1$, $K4 = 1$, and $K5 = 0$, the contents of the scan cells would be "1111000001" (Figure \ref{fig:scan_attack2}). We observe that the same input scan vector upload different values in the scan cells under different key setups. We have also observed in the previous section that no information can be extracted from the scan out response. As an attacker does not know the correct keys of an activated IC, it is not possible to use the scan-chain to read/write the values of all flip-flops in the design. Restricted controllability and observability of the scan chain inhibits an attacker to use the inputs and outputs of the flip-flops as pseudo primary outputs and inputs, respectively. Thus, our proposed encryption strategy has the inherent ability to prevent SAT-based attack. Unlike other SAT-preventive measures\cite{yasin2016sarlock, xiemitigating}, \textit{Encrypt Flip-Flop} strategy does not require any extra hardware infrastructure to rule out SAT-based attack. Moreover, removal attack is also not applicable on the proposed method.

		\section{Experimental Results}
		In this section, we present the results of our experiments on several ISCAS'89 and ITC'99 benchmarks \cite{albrecht2005iwls}. We select the benchmarks of different sizes and encrypt them using our proposed \textit{Encrypt Flip-flip} strategy. The key-gate location selection algorithm (Algorithm \ref{algo:key-location}) is implemented using a C code which identifies the flip-flops affecting the largest overlapping input cone of dependency of the outputs of a circuit. All such flip-flops are the potential candidates for encryption. We select $K$ random flip-flops from these flip-flops and encrypt them by inserting a MUX in front of each of them. Table \ref{tab:circuit_details} reports the number of inputs, outputs, logic gates and flip-flops present in each of the selected benchmarks. The benchmarks $s5378$, $s9234$, $s13207$, and $s15850$ are relatively small, with gate counts in the range of 10K. $s38417$, $s38584$, $b17$, $b20$, $b21$, and $b22$ are the medium size benchmarks, with gate counts less than 30K, while $b18$ and $b19$ are large benchmarks, with gate counts 100K and 200K respectively. Columns 6 and 7 of the table report the number of flip-flops present in the largest overlapping input cone of dependency ($ICOD_{overlap}$) and the number of outputs getting affected by this $ICOD_{overlap}$, respectively. For example, 72 out of 152 outputs of the benchmark $s13207$ form the largest $ICOD_{overlap}$, which includes 377 (out of 638) flip-flops. Therefore, any subset of these 377 flip-flops can be encrypted using our proposed method. One may observe that all the benchmarks include a sufficient number of flip-flops in the largest $ICOD_{overlap}$, which allows us to encrypt a design with an adequate number of keys. A wrong key can corrupt only the outputs which are included in the $ICOD_{overlap}$. Rest of the outputs remain unaffected irrespective of the correctness of the keys, as the values of the flip-flops included in the $ICOD_{overlap}$ do not propagate to those outputs. Column 8 of the table reports the percentage of the total number of outputs covered by the largest $ICOD_{overlap}$ of each of the benchmarks. The affected output coverage of the benchmarks $s5378$, $s38584$, $b18$, $b19$, $b20$, $b21$, and $b22$ is very high, which indicates that most of the outputs of these circuits get affected by any wrong key. On the other hand, the affected output coverage of the benchmarks $s9234$ and $s13207$ are medium, while the $ICOD_{overlap}$ of the benchmarks $s15850$, $s38417$, and $b17$ have poor output coverages. Thus, only a few outputs of these benchmarks get affected by a wrong key. Column 9 of the table reports the encryption time of our proposed \textit{Encrypt Flip-Flop} technique, which is performed on a computer with 3.20 GHz Intel(R) Core(TM) i5-3470 processor and 4 GB RAM. Our proposed technique takes very small amount of time (less than a minute) to encrypt the smaller benchmarks like $s5378$, $s9234$, $s13207$, and $s15850$. The medium size benchmarks like $s38417$, $b20$, $b21$, and $b22$ can be encrypted in less than 10 minutes using our proposed method, while the other two medium size benchmarks $s38584$ and $b17$ require less than 45 minutes to get encrypted. The encryption time of the larger benchmark $b18$ (100K gates) is less than 2 hours. Only the benchmark $b19$, with 200K gates, required more than 11 hours to get encrypted by our proposed method. Therefore, on an average, the encryption time of our proposed strategy is reasonably low, which indicates the simplicity of the \textit{Encrypt Flip-flip} technique.
		
		\begin{figure*}[!ht]
			\centering
			\captionsetup{justification=centering}
			\begin{subfigure}[t]{0.33\textwidth}
				\centering
				\includegraphics[width=\textwidth]{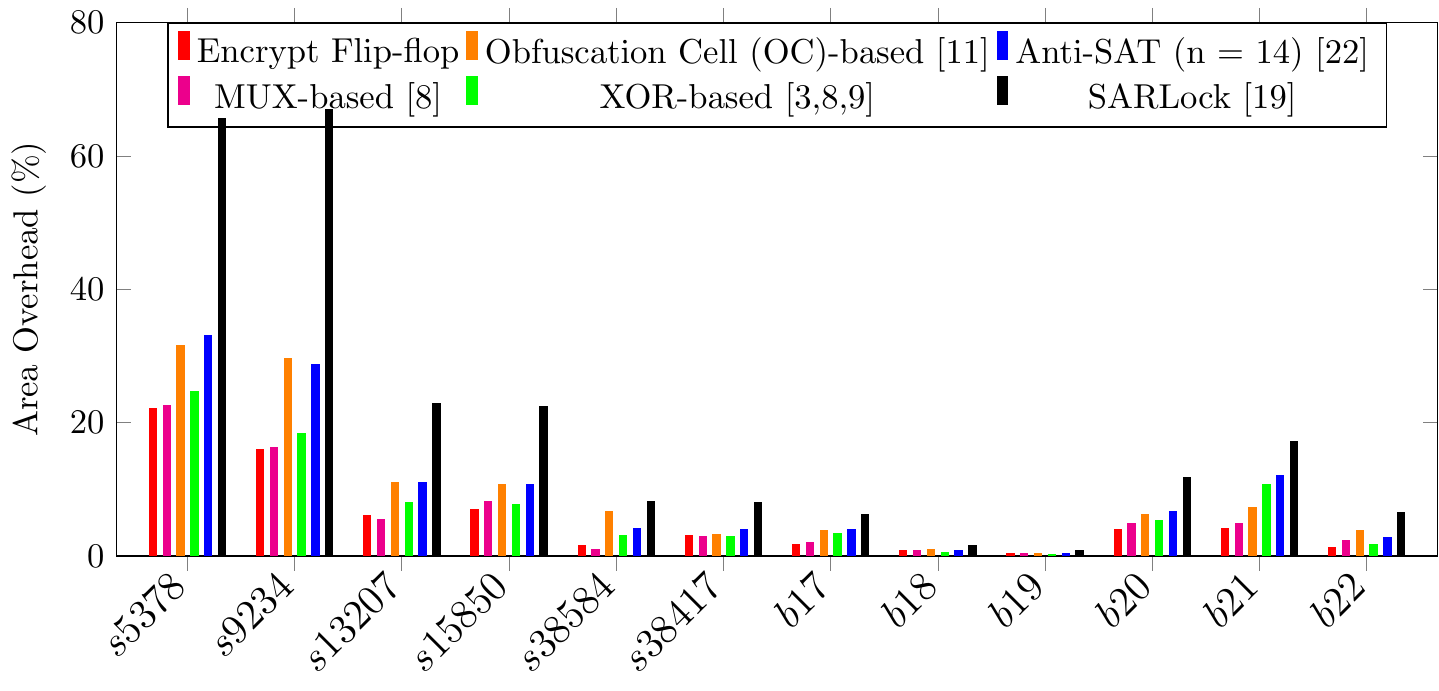}
				\caption{Area Overhead}
				\label{fig:area128}
				\vspace{-0.4cm}
			\end{subfigure}%
			\begin{subfigure}[t]{0.33\textwidth}
				\centering
				\includegraphics[width=\textwidth]{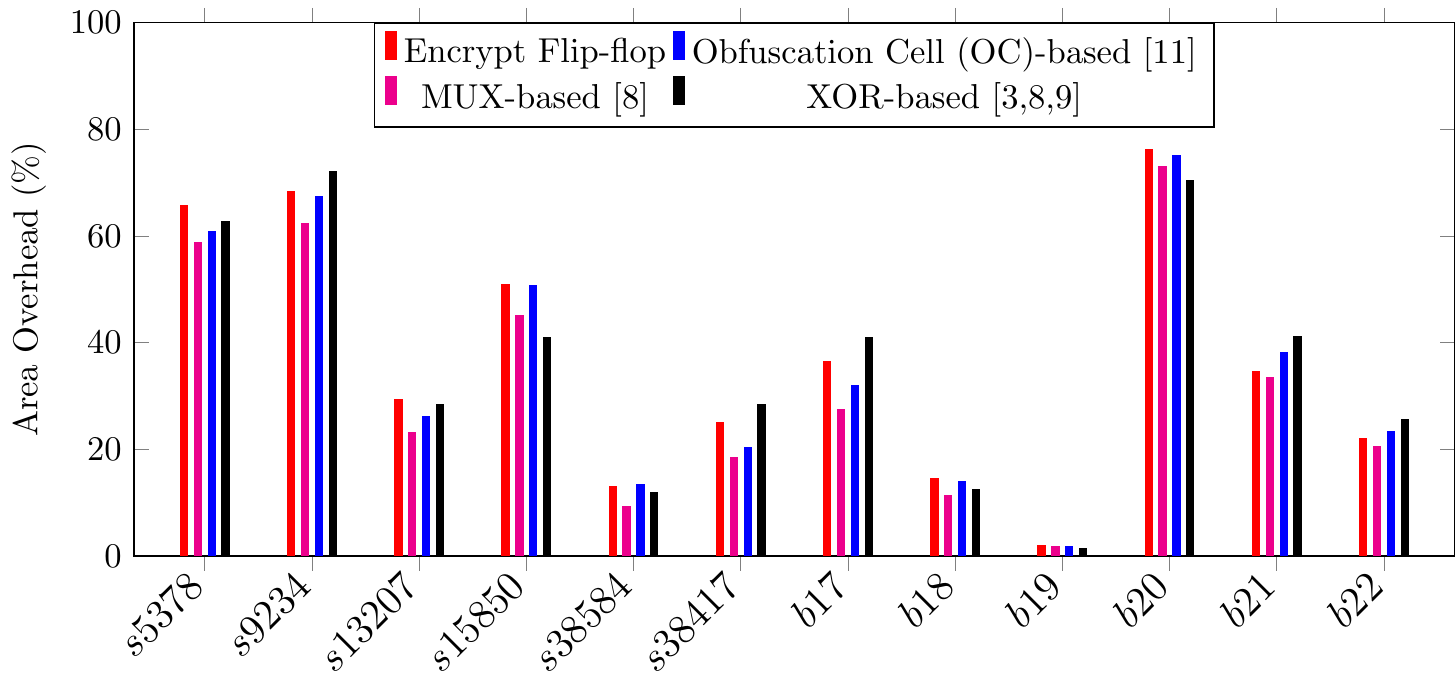}
				\caption{Power Overhead}
				\label{fig:power128}
				\vspace{-0.4cm}
			\end{subfigure}%
			\begin{subfigure}[t]{0.33\textwidth}
				\centering
				\includegraphics[width=\textwidth]{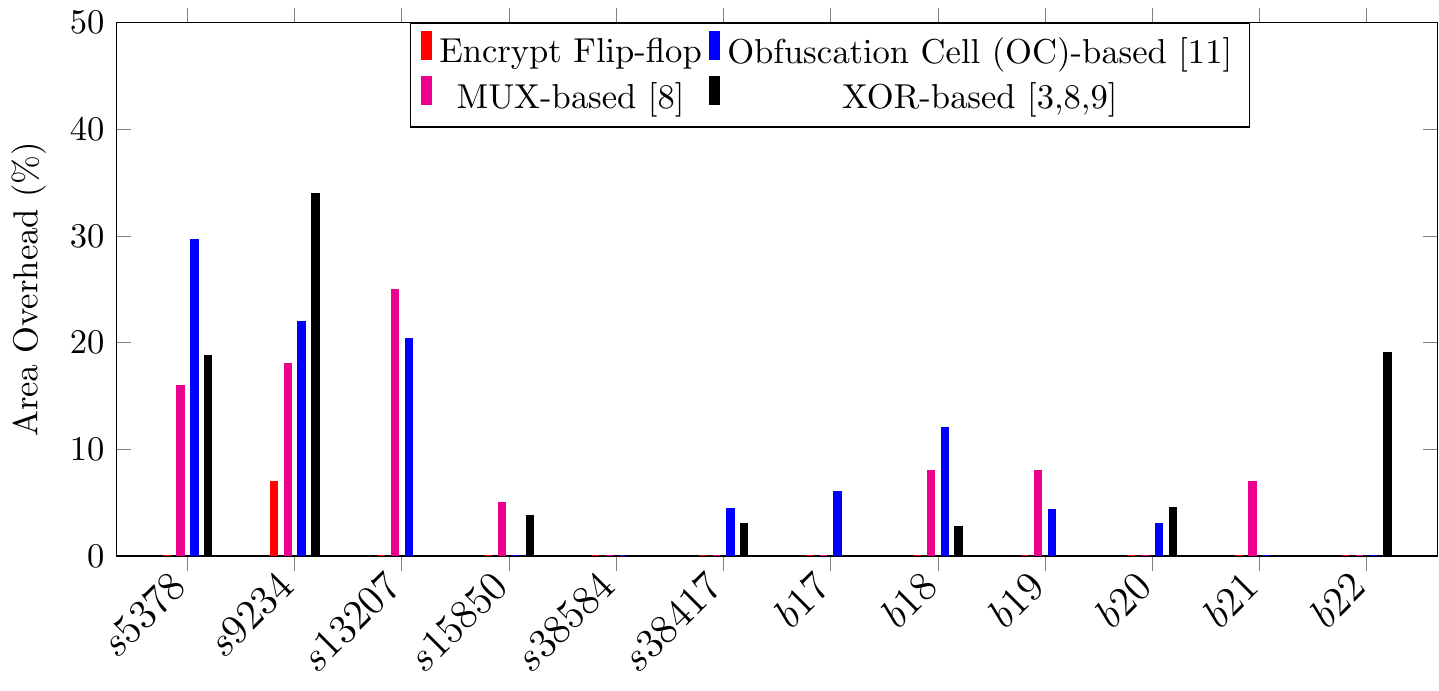}
				\caption{Delay Overhead}
				\label{fig:delay128}
				\vspace{-0.4cm}
			\end{subfigure}
			\caption{Comparison of area, power and delay overheads between different logic encryption strategies ($K = 123$ for $s9234$, $K = 128$ for others)}
			\label{fig:overhead}
			\vspace{-1cm}
		\end{figure*}

		\vspace{-0.5cm}
		
		\subsection{Area, Power, and Delay Overheads}
		To evaluate the area, power, and delay overheads of our proposed encryption technique, we encrypt the benchmarks with 128-bit keys.
		We synthesize each of the designs (both encrypted and unencrypted versions) using Synopsys Design Vision tool \cite{Synopsys1} (using Faraday 90nm library), and calculate the overheads of our proposed scheme. To compare the overheads with other encryption strategies, we also encrypt the benchmarks using typical XOR/XNOR based (\cite{rajendran2015fault,roy2008epic,yasin2015improving}), MUX-based \cite{rajendran2015fault} and Obfuscation Cell (OC) based \cite{zhang2016practical} encryption strategies and synthesize those encrypted benchmarks and calculate the overheads for each of them. Figure \ref{fig:overhead} compares the area, power, and delay overheads of our \textit{Encrypt Flip-flop} method with other encryption strategies. Please note that the benchmark $s9234$ has a maximum of 123 flip-flops as potential candidate for encryption (refer to Table \ref{tab:circuit_details}). Therefore, the results shown in the Figure \ref{fig:overhead} consider $K = 123$ for $s9234$ and $K = 128$ for rest of the benchmarks.

		Figure \ref{fig:area128} also reports the estimated area overheads of SARLock \cite{yasin2016sarlock} and Anti-SAT \cite{xiemitigating} methods, which are capable of preventing the powerful SAT attack. For a $K$-bit key, $K+1$ XOR gates and $2K+1$ AND gates are required to build the infrastructure of the SARLock method \cite{yasin2016sarlock}, while the extra hardware requirements for the Anti-SAT method (considering $N$-bit Anti-SAT block) are $3N+1$ XOR/XNOR gates, $N$ 2-input MUXes, 1 $N$-input NAND gate, 1 $N$-input AND gate and 1 2-input AND gate \cite{xiemitigating}. Both of these SARLock and Anti-SAT methods need to be integrated with any typical XOR/XNOR-based (either Strong Logic Locking or Fault Analysis) encryption technique to encrypt a design. Due to the high implementation complexity, we have not implemented the SARLock and Anti-SAT methods. Instead, we estimate the area overheads of these two methods by adding the area of the extra hardware incurred by them with the hardware overhead of the XOR/XNOR based encryption technique. For this purpose, we refer to the data sheet of Faraday 90nm library \cite{online3}. Table \ref{tab:cell_area} presents the area units of different logic cells (from the data sheet of Faraday 90nm library) that we consider while estimating the area of the extra hardware incurred by the SARLock and Anti-SAT methods. The method gives a rough estimation of the area overheads of the SARLock and Anti-SAT methods.  
		
		\vspace{-0.4cm}
		\begin{table}[!ht]
			\centering
			\caption{Cell area of different logic cells reported in Faraday 90nm library \cite{online3}}
			\label{tab:cell_area}
			\begin{tabular}{|c|c|}
				\hline
				\textbf{Cell Name} & \textbf{Area Unit} \\ \hline
				2 input XOR & 10 \\ \hline
				2 input XNOR & 10 \\ \hline
				2 input MUX & 9 \\ \hline
				2 input AND & 5 \\ \hline
				2 input NAND & 4 \\ \hline
			\end{tabular}
			\vspace{-0.3cm}
		\end{table}
		
		We can observe from Figure \ref{fig:area128} that the area overhead of the proposed \textit{Encrypt Flip-Flop} method is comparable with MUX-based \cite{rajendran2015fault}, OC-based \cite{zhang2016practical}, and XOR/XNOR based (\cite{rajendran2015fault,roy2008epic,yasin2015improving}) encryption strategies. However, unlike \textit{Encrypt Flip-Flop} technique, none of these methods can prevent SAT attack. On the other hand, SAT-resilient techniques like SARLock \cite{yasin2016sarlock} and Anti-SAT \cite{xiemitigating} have higher area overheads compared to our proposed method. We can also observe that SARLock has higher hardware overheads compared to Anti-SAT. As we have not implemented the SARLock and Anti-SAT techniques, we could not calculate the power and delay overheads of these two techniques. Thus, we could not compare the power and delay overheads of our method with these two. Figure \ref{fig:power128} shows the comparison of power overhead our method with MUX-based, OC-based, and XOR/XNOR based methods. Here also we can observe that the power overhead of our \textit{Encrypt Flip-Flop} method is comparable with other encryption techniques. Figure \ref{fig:delay128} shows the comparison of delay overheads of different strategies. Presence of key-gates in the critical path of a circuit increases its delay. Thus, to encrypt a design without its performance degradation, our strategy avoids the encryption of the flip-flops which are present in the critical path of a circuit. The delay overhead of our \textit{Encrypt Flip-Flop} technique is zero for all the benchmarks except $s9234$, when it is encrypted with 123-bit key. This is because we encrypt all the candidate flip-flops, some of which are also a part of the critical paths of the circuit. Excluding those flip-flops from encryption ensures zero delay overheads for this circuit as well.
		
		\begin{figure*}[!ht]
			\centering
			\captionsetup{justification=centering}
			\begin{subfigure}[t]{0.24\textwidth}
				\centering
				\includegraphics[width=\textwidth]{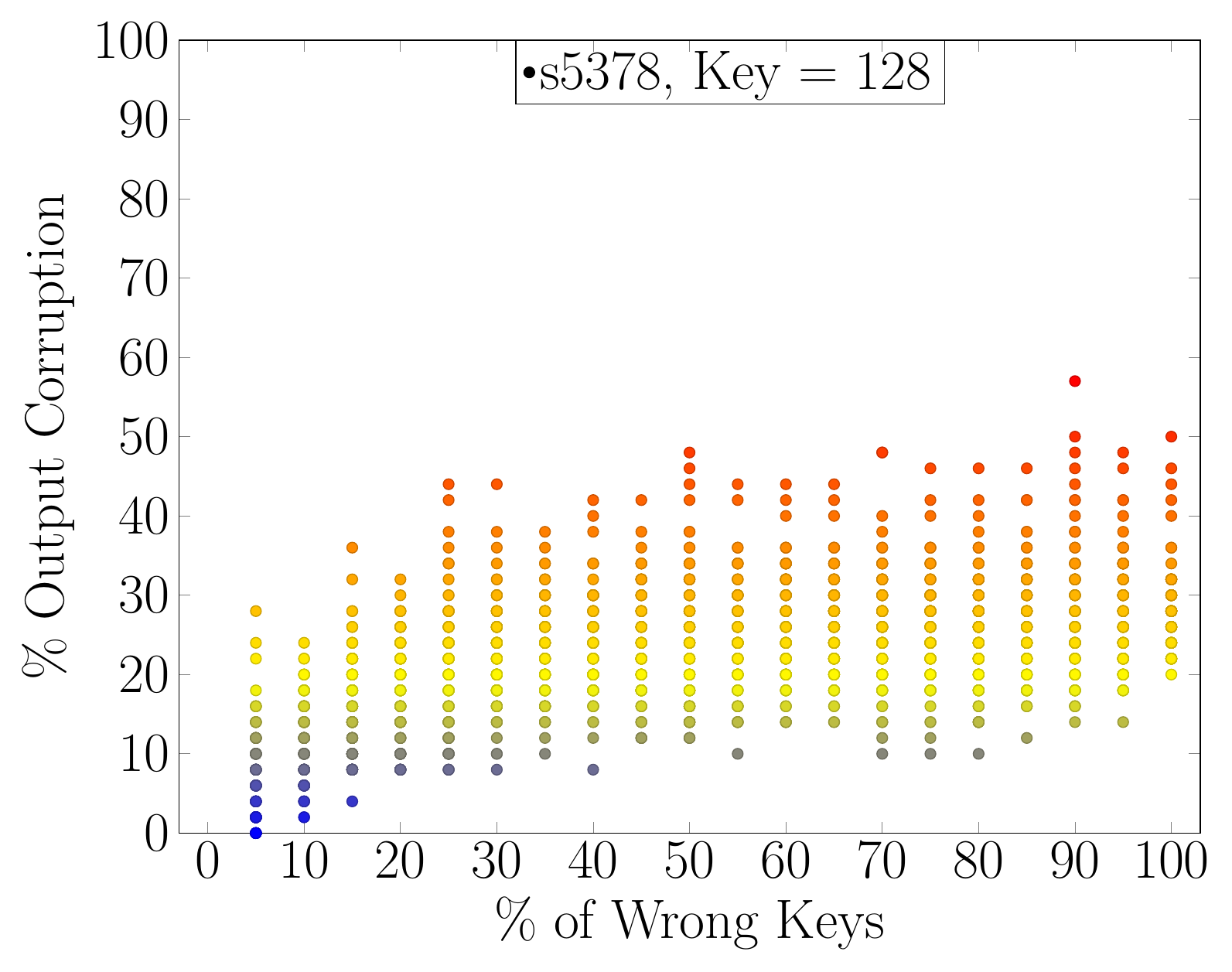}
				\caption{$s5378$}
				\label{fig:s5378}
				\vspace{-0.3cm}
			\end{subfigure}%
			\begin{subfigure}[t]{0.24\textwidth}
				\centering
				\includegraphics[width=\textwidth]{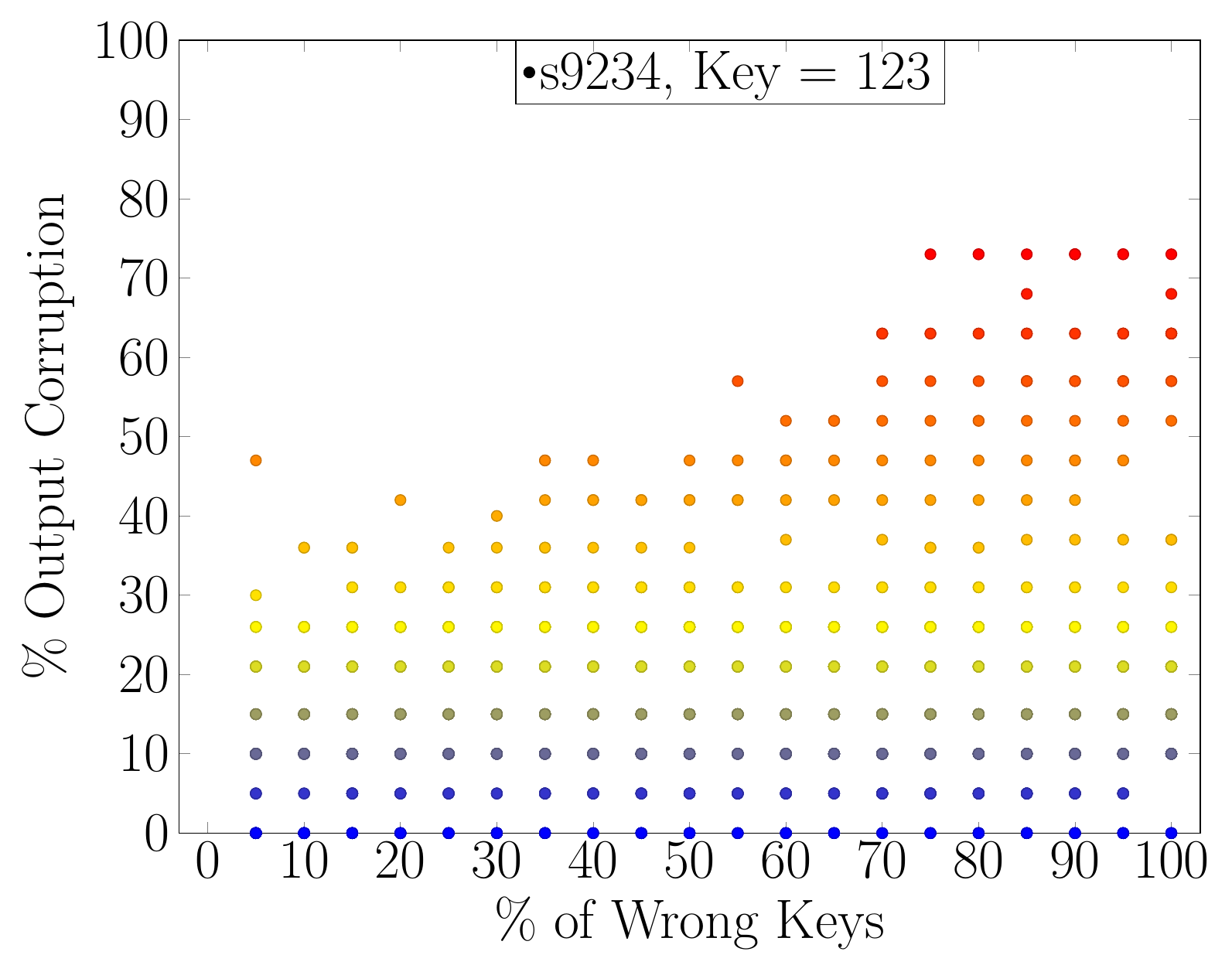}
				\caption{$s9234$}
				\label{fig:s9234}
				\vspace{-0.3cm}
			\end{subfigure}%
			\begin{subfigure}[t]{0.24\textwidth}
				\centering
				\includegraphics[width=\textwidth]{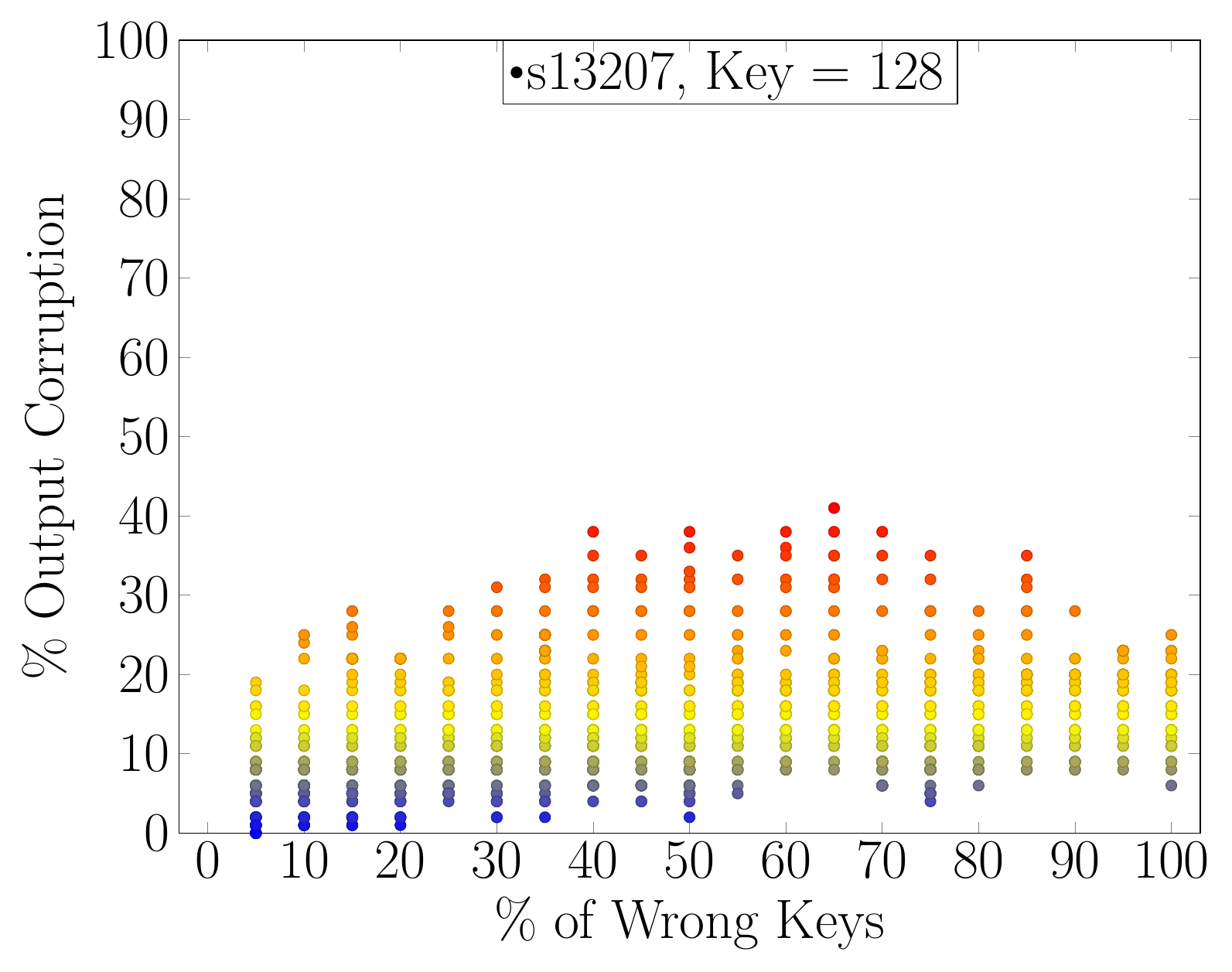}
				\caption{$s13207$}
				\label{fig:s13207}
				\vspace{-0.3cm}
			\end{subfigure}%
			\begin{subfigure}[t]{0.24\textwidth}
				\centering
				\includegraphics[width=\textwidth]{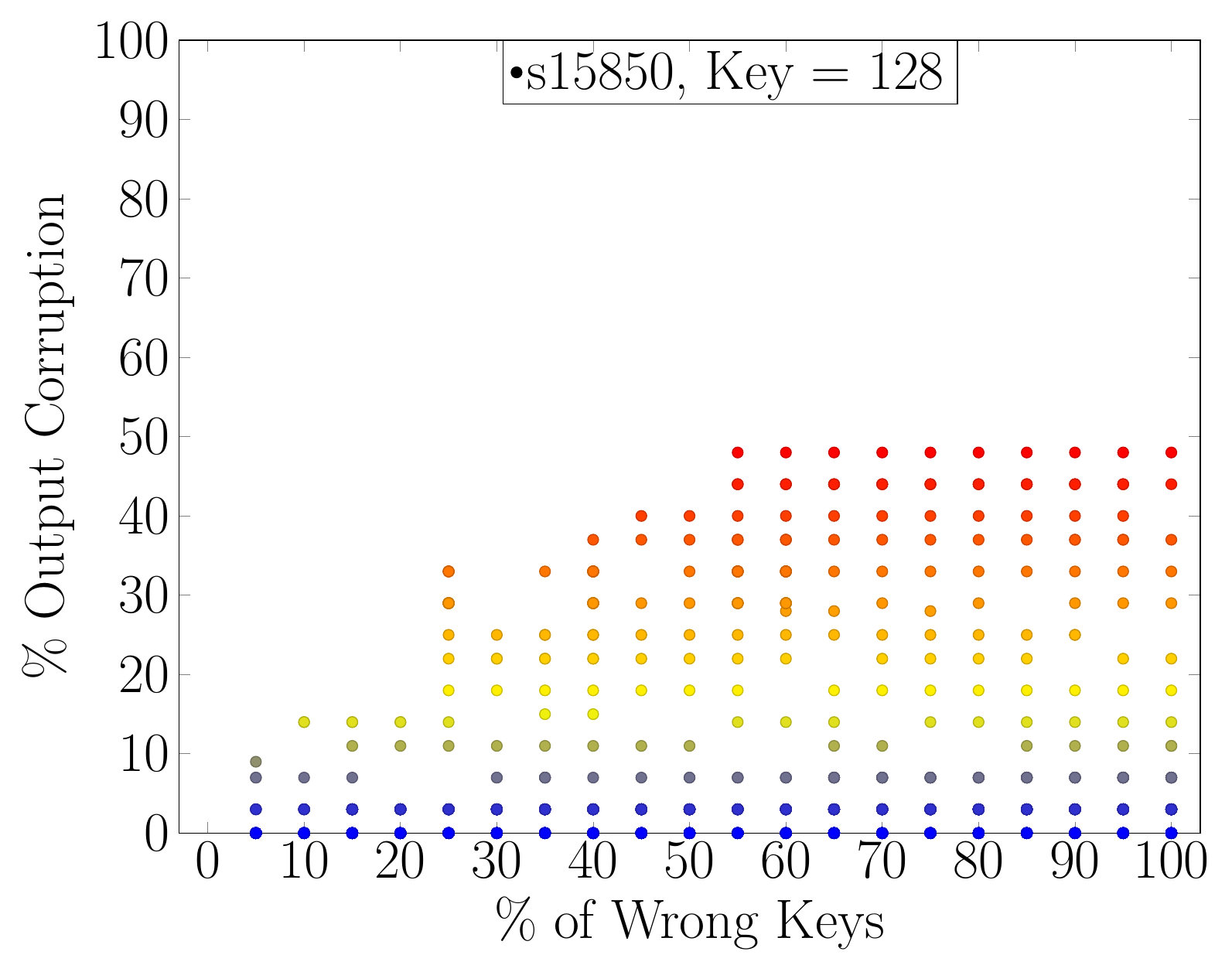}
				\caption{$s15850$}
				\label{fig:s15850}
				\vspace{-0.3cm}
			\end{subfigure}
			\begin{subfigure}[t]{0.24\textwidth}
				\centering
				\includegraphics[width=\textwidth]{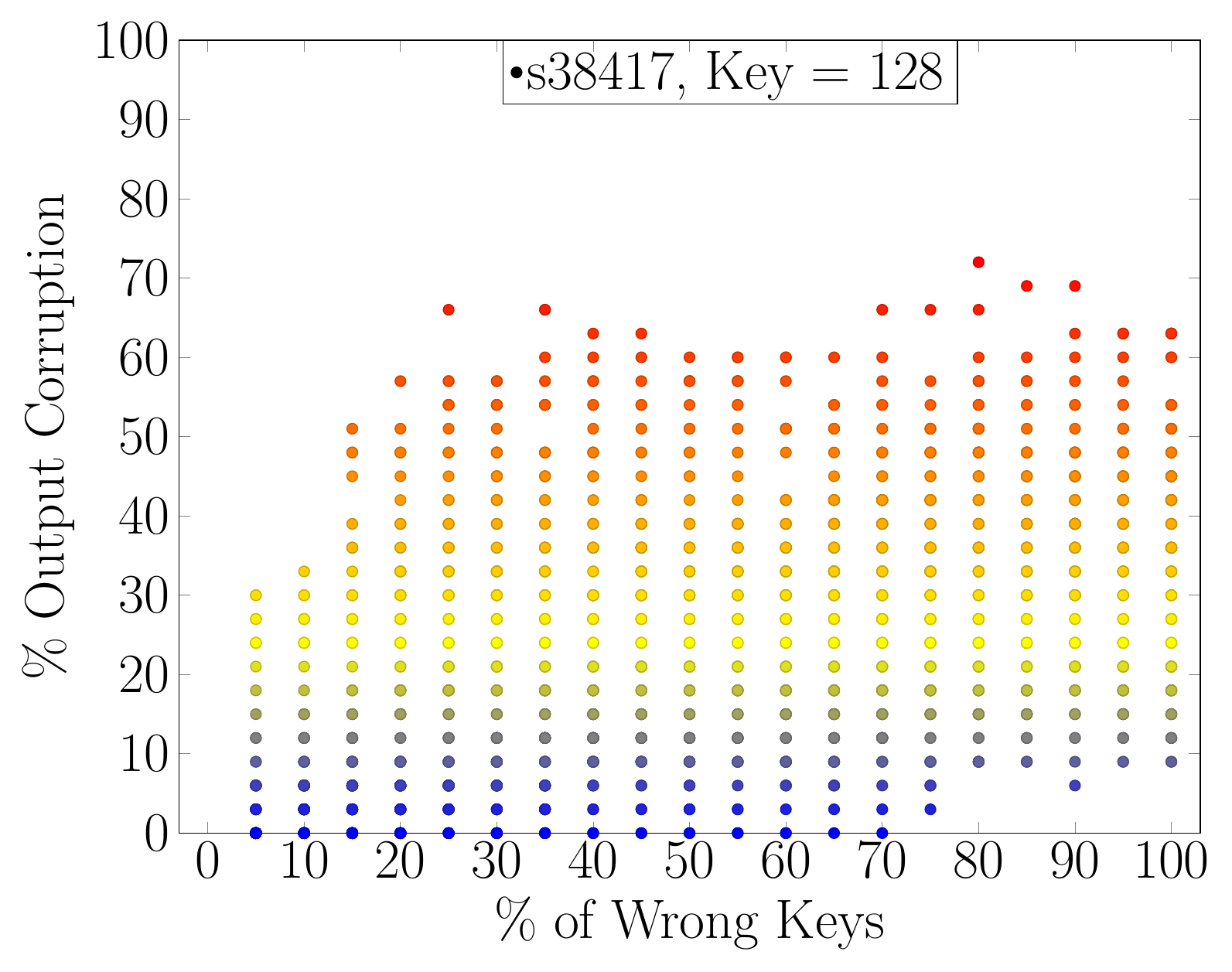}
				\caption{$s38417$}
				\label{fig:s38417}
				\vspace{-0.3cm}
			\end{subfigure}%
			\begin{subfigure}[t]{0.24\textwidth}
				\centering
				\includegraphics[width=\textwidth]{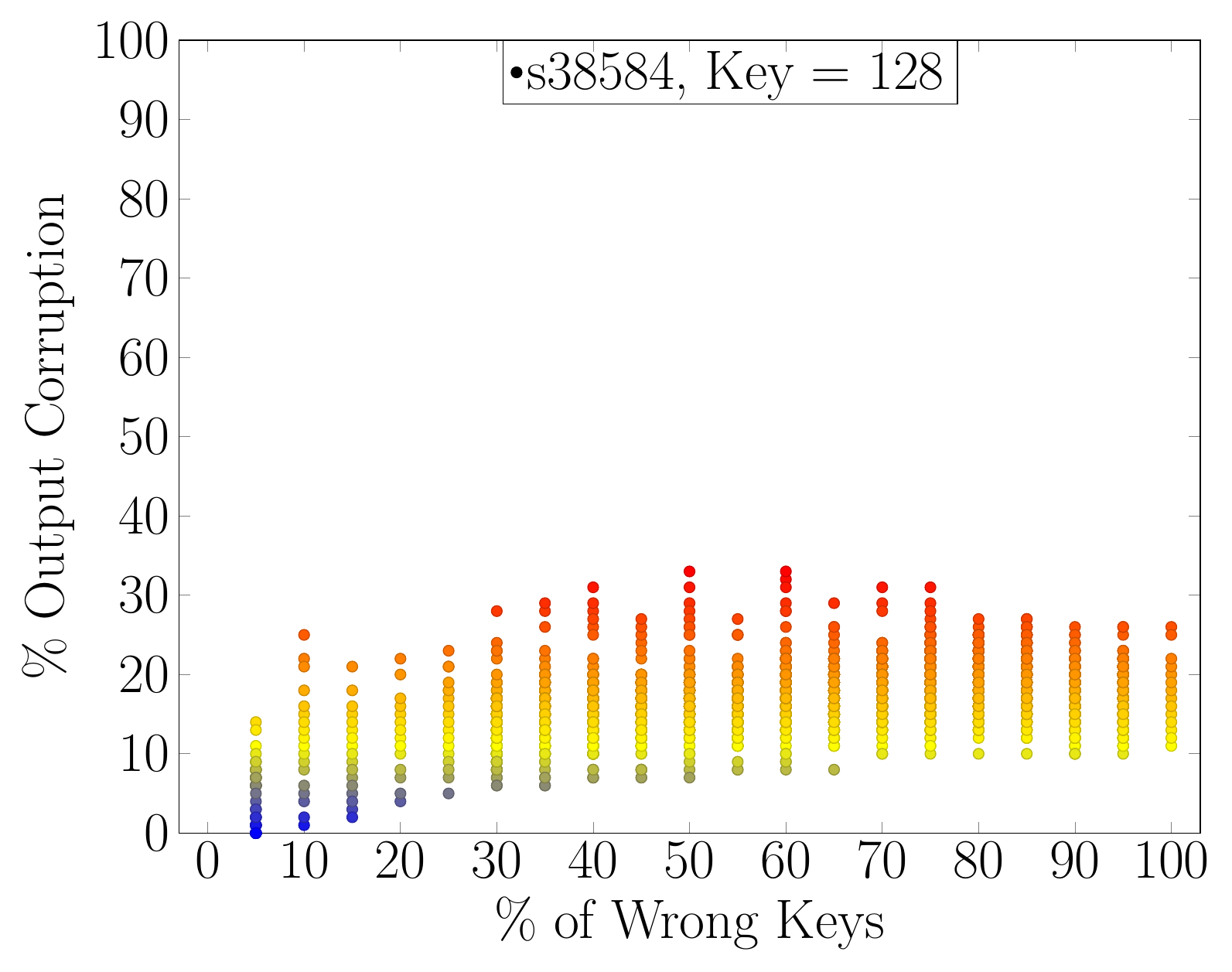}
				\caption{$s38584$}
				\label{fig:s38584}
				\vspace{-0.3cm}
			\end{subfigure}%
			\begin{subfigure}[t]{0.24\textwidth}
				\centering
				\includegraphics[width=\textwidth]{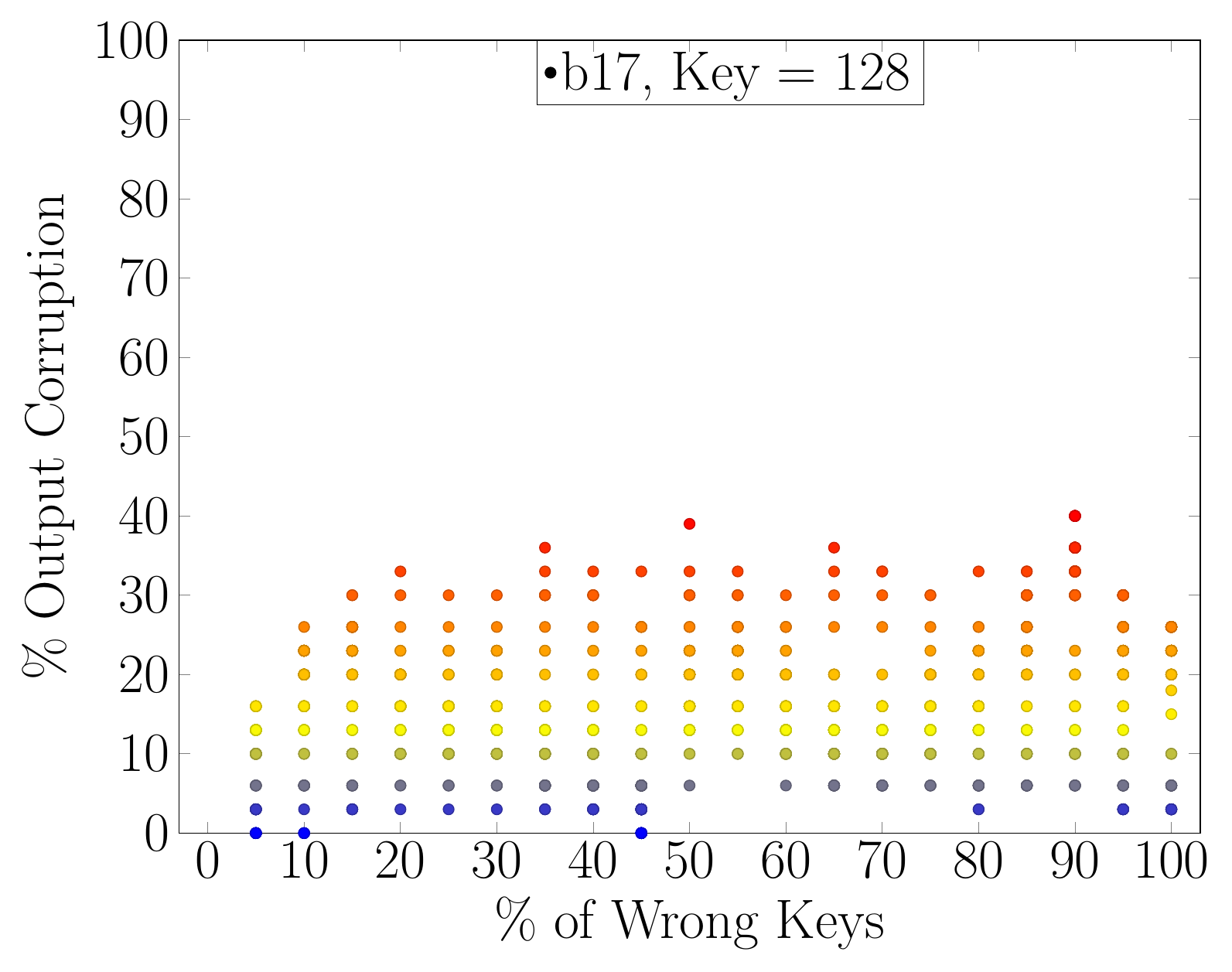}
				\caption{$b17$}
				\label{fig:b17}
				\vspace{-0.3cm}
			\end{subfigure}%
			\begin{subfigure}[t]{0.24\textwidth}
				\centering
				\includegraphics[width=\textwidth]{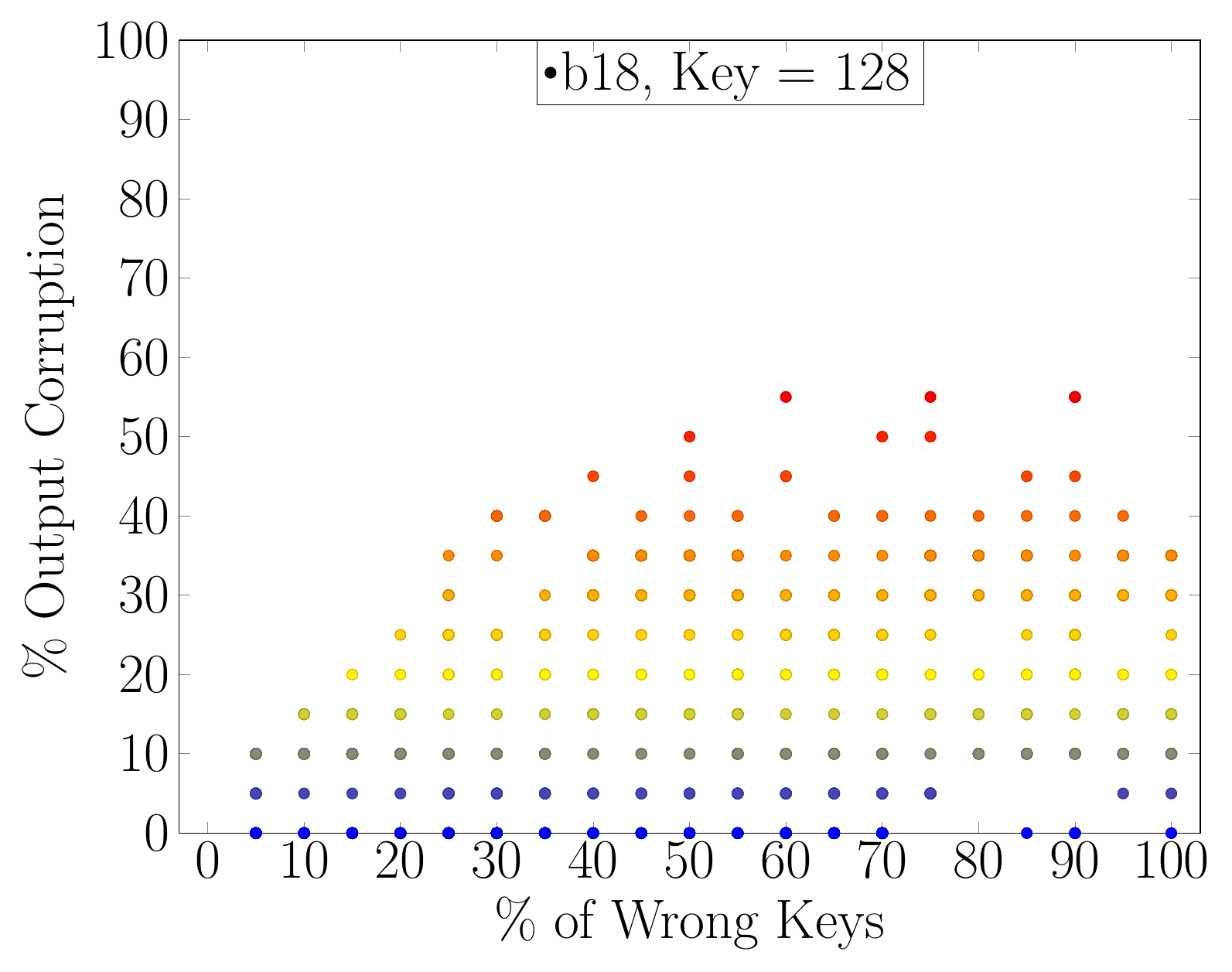}
				\caption{$b18$}
				\label{fig:b18}
				\vspace{-0.3cm}
			\end{subfigure}
			\begin{subfigure}[t]{0.24\textwidth}
				\centering
				\includegraphics[width=\textwidth]{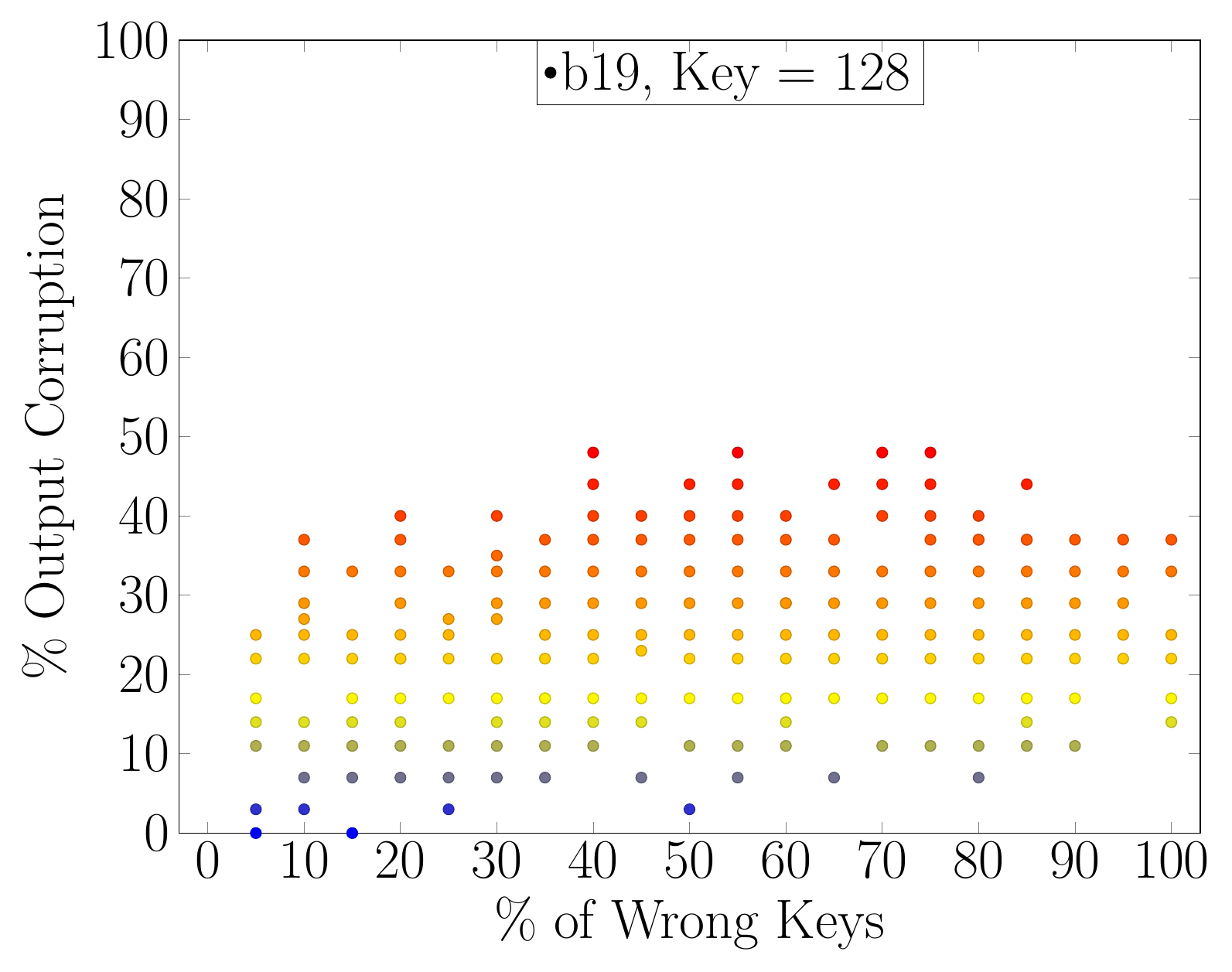}
				\caption{$b19$}
				\label{fig:b19}
				\vspace{-0.3cm}
			\end{subfigure}%
			\begin{subfigure}[t]{0.24\textwidth}
				\centering
				\includegraphics[width=\textwidth]{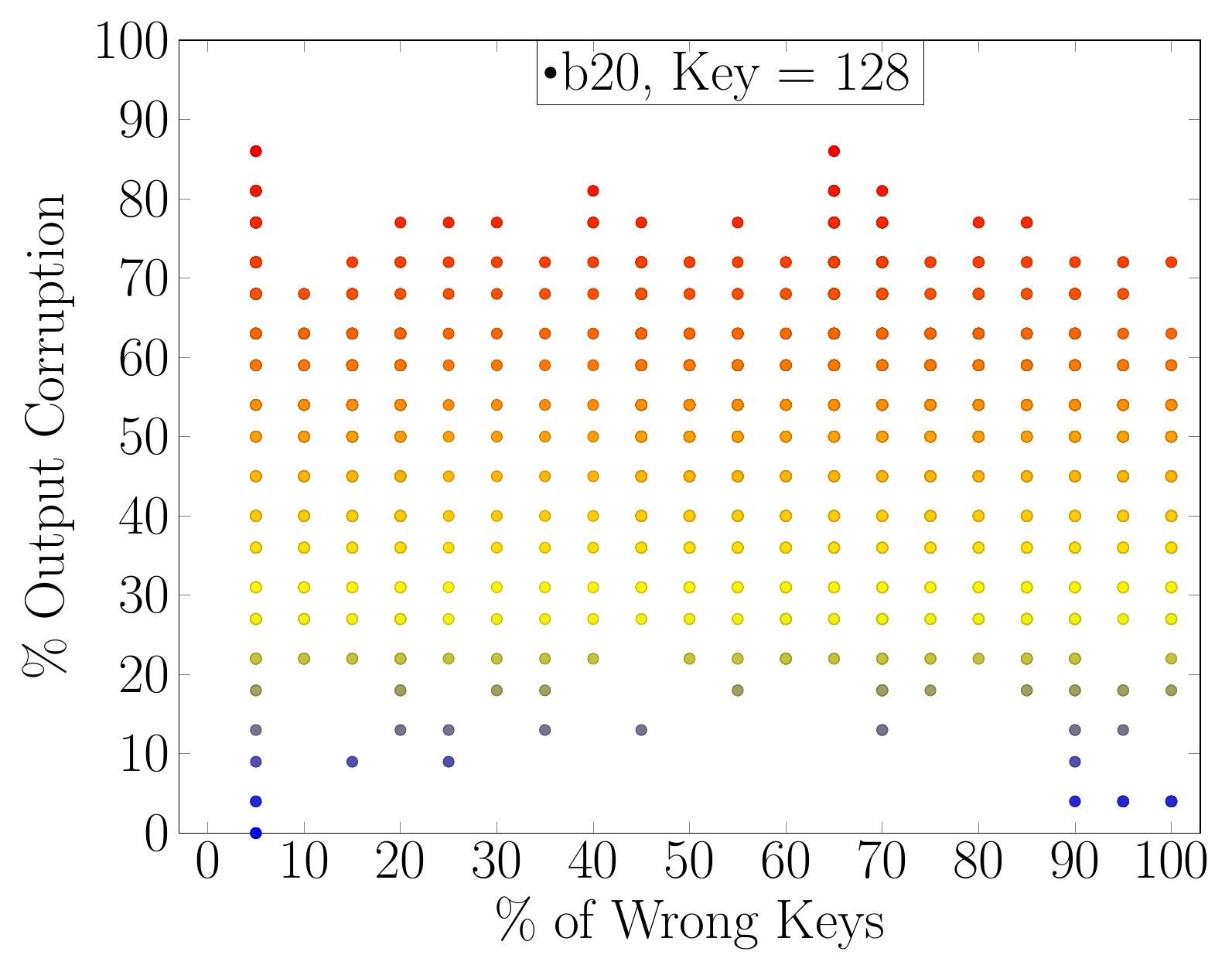}
				\caption{$b20$}
				\label{fig:b20}
				\vspace{-0.3cm}
			\end{subfigure}%
			\begin{subfigure}[t]{0.24\textwidth}
				\centering
				\includegraphics[width=\textwidth]{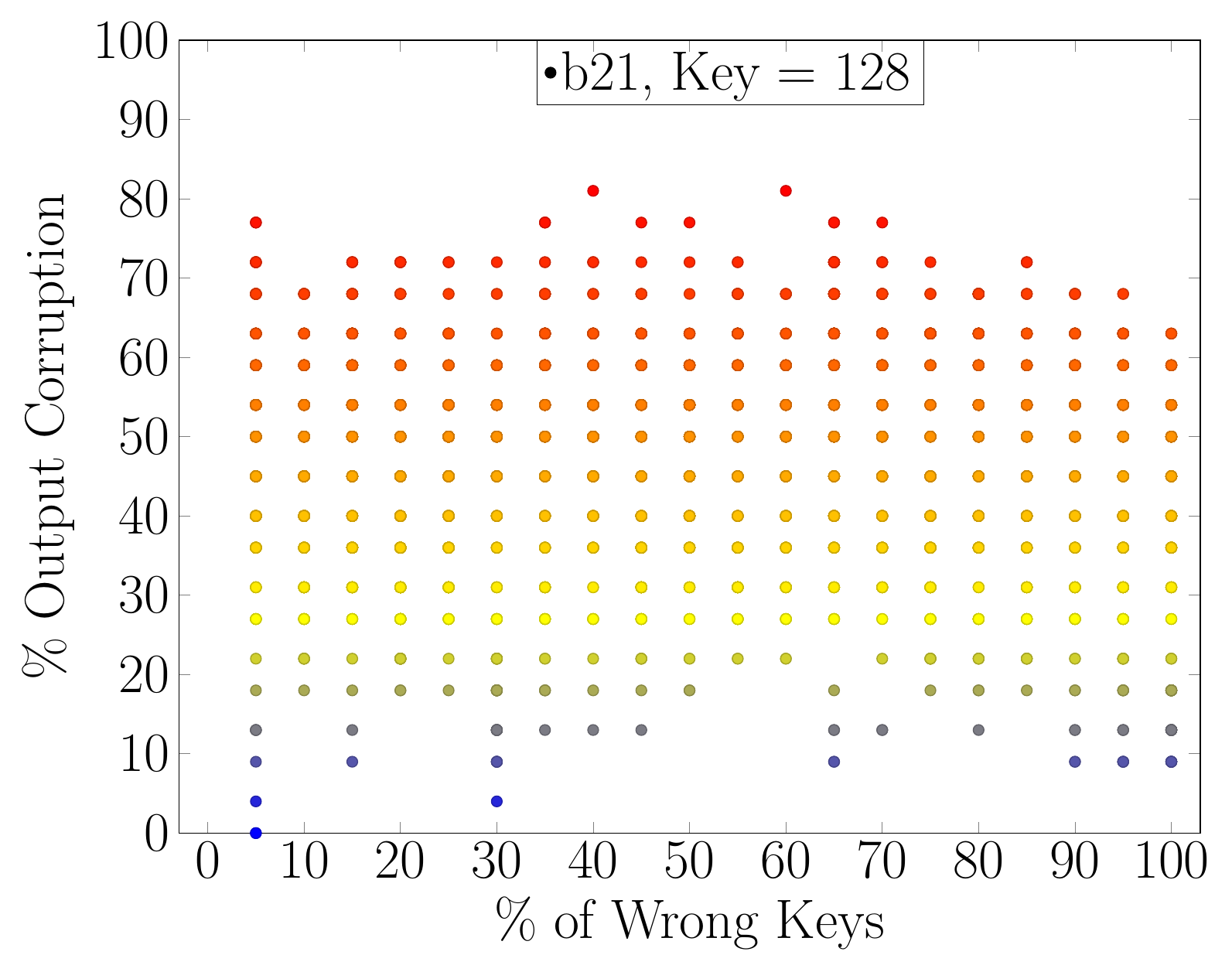}
				\caption{$b21$}
				\label{fig:b21}
				\vspace{-0.3cm}
			\end{subfigure}%
			\begin{subfigure}[t]{0.24\textwidth}
				\centering
				\includegraphics[width=\textwidth]{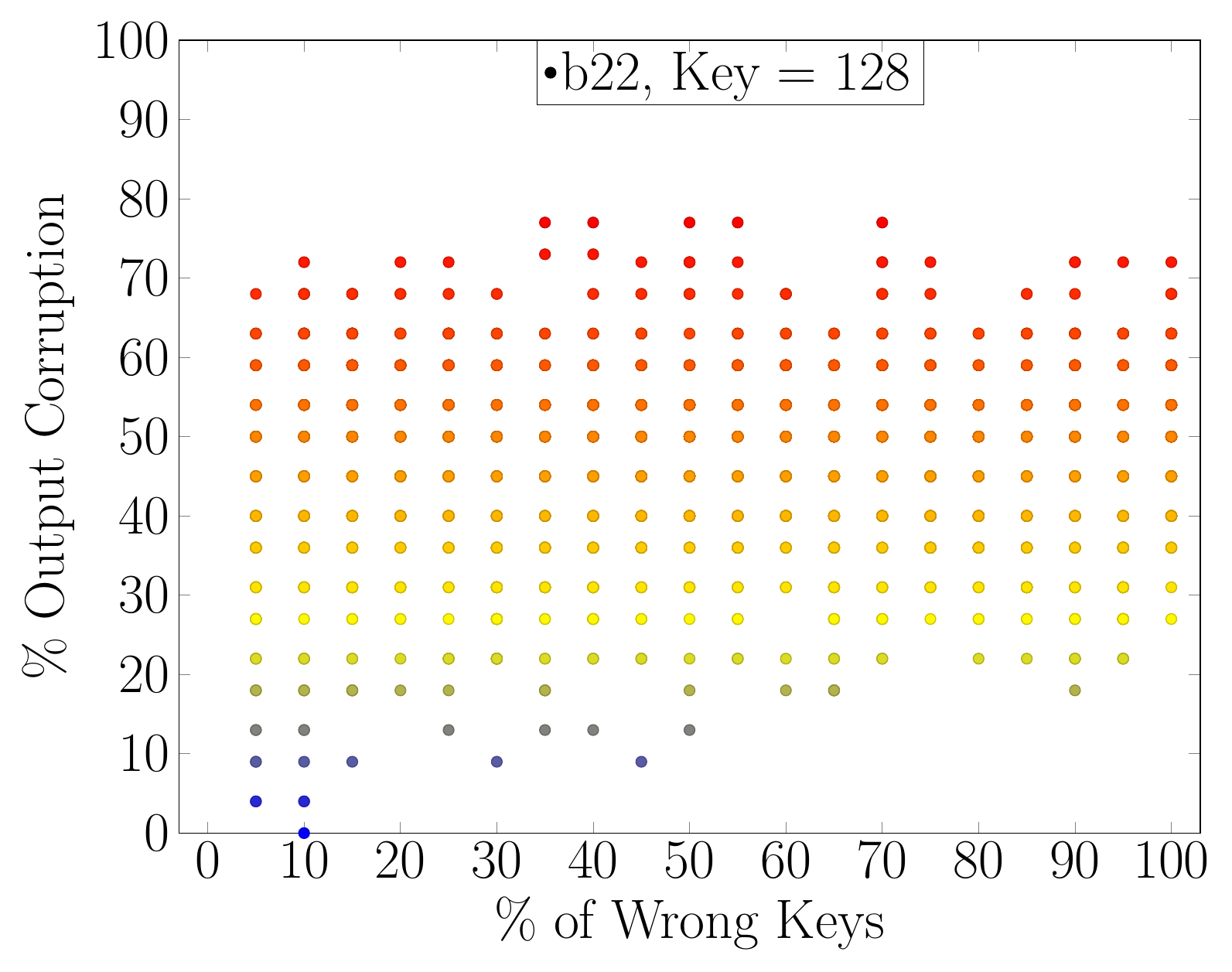}
				\caption{$b22$}
				\label{fig:b22}
				\vspace{-0.3cm}
			\end{subfigure}
			\caption{Variation of \% output corruption with the variation of \% of wrong keys for different benchmarks ($K = 128$)}
			\label{fig:output_corruption}
			\vspace{-0.7cm}
		\end{figure*}

		\vspace{-0.3cm}
		\subsection{Output Corruptibility For Wrong Keys}
		One important metric to measure the quality of any encryption technique is its ability to corrupt the outputs for any wrong key. To measure the output corruptibility of our proposed method, we simulate each benchmark with 10000 random input patterns. We vary the percentage of wrong keys and measure the Hamming Distance between the correct and the obtained outputs. Figure \ref{fig:output_corruption} shows the variation of \% output corruption with the variation of \% of wrong keys for different benchmarks (for $K = 128$). Please note that the data shown in the Figure \ref{fig:output_corruption} consider the outputs affected by the $ICOD_{overlap}$ (refer to Table \ref{tab:circuit_details}) while calculating the \% output corruption. We gradually increase the \% of wrong keys from 5\% to 100\% and check the \% of output corruption for different random input patterns. We can observe from the figure that some benchmarks like $s9234$, $s38417$, $b20$, $b21$, and $b22$ offer high output corruption for wrong keys. Other benchmarks like $s5378$, $s15850$, $b17$, $b18$, and $b19$ offer reasonable output corruption for wrong keys, while the \% output corruptibility of the benchmarks $s13207$ and $s38584$ are low. It can be observed from Table \ref{tab:circuit_details} that the numbers of affected outputs of  $s13207$ and $s38584$ are high. Therefore, even a sufficient number of output bit corruption show low \% output corruption for these two benchmarks. We can also observe from Figure \ref{fig:output_corruption} that for all the benchmarks, we get a zero output corruption for some of the input patterns even for a wrong key. This is because the effect of wrong keys does not propagate to the outputs for those input patterns. However, we found a zero output corruption only for a few input patterns for all the benchmarks.
		
		\begin{figure}[!ht]
			\centering
			\includegraphics[scale = 0.45]{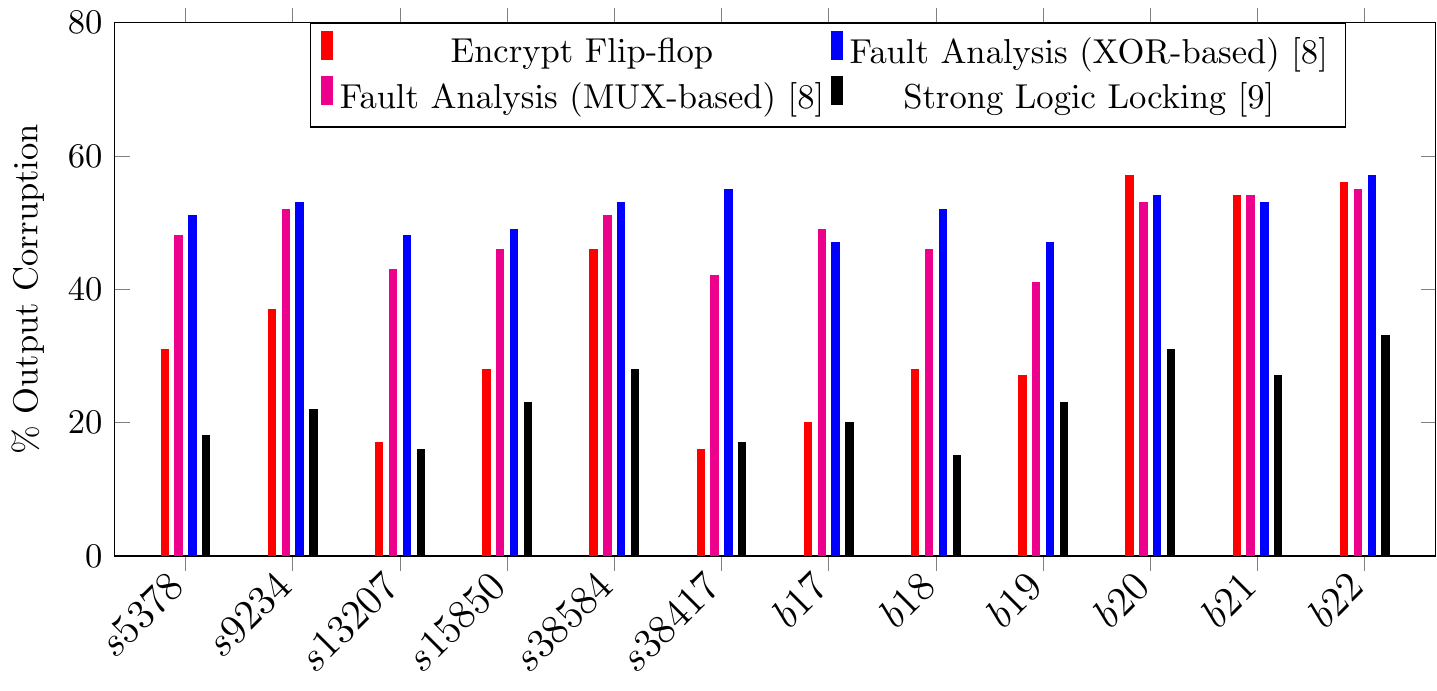}
			\caption{Comparison of average output corruption between different logic encryption strategies for K = 128}
			\label{fig:avg_output_corruption}
			\vspace{-0.3cm}
		\end{figure}

		Figure \ref{fig:avg_output_corruption} shows a comparison of average output corruption between different logic encryption strategies for K = 128. For this comparison, we mainly consider strong logic locking (SLL) \cite{yasin2015improving} and fault analysis (FA) based (both XOR and MUX-based)\cite{rajendran2015fault} strategies.  As Obfuscation Cell \cite{zhang2016practical} and logic cone prevention \cite{lee2015improving} based approaches need to be integrated with either FA or SLL based approaches, we do not consider them in the comparison. We can observe from Figure \ref{fig:avg_output_corruption} that both XOR and MUX-based fault analysis approaches \cite{rajendran2015fault} produce high average output corruption for wrong keys. As the proposed $Encrypt$ $flip$-$Flop$ method does not take any explicit measure to increase output corruption, the average output corruption for this method varies from circuit to circuit. Although the output corruption of the proposed method is not as good as the fault analysis based approach, still it could produce reasonable output corruption for wrong keys. Output corruption for strong logic locking is comparatively low, which is one of the shortcomings of the method.

		\def\checkmark{\tikz\fill[scale=0.25](0,.35) -- (.25,0) -- (1,.7) -- (.25,.15) -- cycle;} 
		
		\begin{table*}[!ht]
			\centering
			\caption{A comparative study between different logic encryption strategies}
			\label{tab:final_comparison}
			\resizebox{\textwidth}{!}{%
				\begin{tabular}{|c|c|c|c|c|l|c|c|c|c|}
					\hline
					\multirow{2}{*}{\textbf{\begin{tabular}[c]{@{}c@{}}Encryption\\ Technique\end{tabular}}} & \multicolumn{5}{c|}{\textbf{Resilience Against Different Attacks}} & \multirow{2}{*}{\textbf{\begin{tabular}[c]{@{}c@{}}Output\\ Corruptability\end{tabular}}} & \multirow{2}{*}{\textbf{\begin{tabular}[c]{@{}c@{}}Hardware\\ overhead\\ (K-bit key)\end{tabular}}} & \multirow{2}{*}{\textbf{\begin{tabular}[c]{@{}c@{}}Encryption\\ time\end{tabular}}} & \multirow{2}{*}{\textbf{\begin{tabular}[c]{@{}c@{}}Implementation\\ Complexity\end{tabular}}} \\ \cline{2-6}
					& \textbf{\begin{tabular}[c]{@{}c@{}}Path\\ Sensitization \cite{yasin2015improving}\end{tabular}} & \textbf{\begin{tabular}[c]{@{}c@{}}Logic\\ Cone \cite{lee2015improving}\end{tabular}} & \textbf{\begin{tabular}[c]{@{}c@{}}Hill\\ Climbing \cite{plaza2015solving}\end{tabular}} & \textbf{SAT \cite{subramanyan2015evaluating}} & \multicolumn{1}{c|}{\textbf{\begin{tabular}[c]{@{}c@{}}Scan\\ Based\end{tabular}}} &  &  &  &  \\ \hline
					\textbf{Random \cite{roy2008epic}} & $\times$ & $\times$ & $\times$ & $\times$ & $\times$ & Low & \begin{tabular}[c]{@{}c@{}}$K$ XOR/\\ (XNOR + NOT)\end{tabular} & Very Fast & Very Simple \\ \hline
					\textbf{\begin{tabular}[c]{@{}c@{}}Fault Analysis (FA) \\ (XOR-based) \cite{rajendran2015fault}\end{tabular}} & $\times$ & $\times$ & $\times$ & $\times$ & $\times$ & High & \begin{tabular}[c]{@{}c@{}}$K$ XOR/\\ (XNOR + NOT)\end{tabular} & Slow & Medium \\ \hline
					\textbf{\begin{tabular}[c]{@{}c@{}}Fault Analysis \\ (MUX-based)\cite{rajendran2015fault}\end{tabular}} & $\times$ & $\times$ & $\times$ & $\times$ & $\times$ & High & $K$ MUX & Slow & Medium \\ \hline
					\textbf{\begin{tabular}[c]{@{}c@{}}Strong Logic \\ Locking (SLL) \cite{yasin2015improving}\end{tabular}} & \checkmark & $\times$ & \checkmark & $\times$ & $\times$ & Low & \begin{tabular}[c]{@{}c@{}}$K$ XOR/\\ (XNOR + NOT)\end{tabular} & Slow & High \\ \hline
					\textbf{\begin{tabular}[c]{@{}c@{}}Obfuscation \\ Cell (OC) \cite{zhang2016practical}\end{tabular}} & $\times$ & $\times$ & $\times$ & $\times$ & $\times$ & Low & \begin{tabular}[c]{@{}c@{}}$K$ MUX +\\  $K$ NOT\end{tabular} & Very Fast & Very Simple \\ \hline
					\textbf{\begin{tabular}[c]{@{}c@{}}Logic Cone \\ Prevention \cite{lee2015improving}\end{tabular}} & $\times$ & \checkmark & $\times$ & $\times$ & $\times$ & Low & \begin{tabular}[c]{@{}c@{}}$K$ (MUX +\\  XOR/XNOR)\end{tabular} & Medium & Medium \\ \hline
					\textbf{\begin{tabular}[c]{@{}c@{}}External \\ Key-Dependency \cite{karmakar2017new}\end{tabular}} & \checkmark & \checkmark & \checkmark & $\times$ & $\times$ & High & \begin{tabular}[c]{@{}c@{}}$4K$ XOR/\\ (XNOR+NOT)\end{tabular} & Slow & High \\ \hline
					\textbf{SLL+SARLock \cite{yasin2016sarlock}}& \checkmark & $\times$ & \checkmark & \checkmark & $\times$ & Low & \begin{tabular}[c]{@{}c@{}}$2K+1$ XORs + \\ $2K+1$ ANDs\end{tabular} & Slow & Very High \\ \hline
					\textbf{Anti-SAT + FA \cite{xiemitigating}} & $\times$ & $\times$ & $\times$ & \checkmark & $\times$ & Medium & \begin{tabular}[c]{@{}c@{}}$K+3N+1$ XOR/XNOR + \\ $N$ 2-input MUX + \\ 1 N-input (NAND + AND) \\ + 1 2-input AND\end{tabular} & Slow & Very High \\ \hline
					\textbf{Encrypt Flip-Flop} & \checkmark & \checkmark & \checkmark & \checkmark & \checkmark & \begin{tabular}[c]{@{}c@{}}Varies from circuit\\  to circuit\end{tabular} & $K$ MUX + 1 XOR + 2 AND + 1 OR + 1 DFF & Fast & Medium \\ \hline
				\end{tabular}%
			}
			\vspace{-0.5cm}
		\end{table*}
		\vspace{-0.5cm}
		\subsection{Security Evaluation Against Hill Climbing Attack}
		Hill climbing attack \cite{plaza2015solving} on conventional logic encryption strategies exploits the linear relationship between the number of wrong keys and output corruption. Due to this linearity, an attacker can converge towards the correct keys by iteratively flipping the key-inputs and reducing the output corruption. However, the nature of the output corruptibility of our encryption strategy (Figure \ref{fig:output_corruption}) depicts that the output corruption does not increase linearly with the increase of the wrong keys. For example, a key with 30\% wrong bits can produce 40\% output corruption, while a key with 70\% wrong bits can produce as low as 10 to 5\% output corruption for the benchmark $b18$. Therefore, an attacker cannot predict the percentage of wrong keys by observing the percentage of output corruption. Even if an attacker obtains less output corruption by flipping a random key-bit, it does not guarantee that the new key has less number of wrong bits. Therefore, an attacker cannot take a decision whether to flip a key-bit in an iteration to reduce the wrong bits of a random key. This phenomenon helps our proposed strategy to thwart the Hill Climbing Attack.

		\section{Discussion}
		In this section, we perform a comparative study between different logic encryption strategies in terms of hardware overhead, output corruptibility, implementation complexity, encryption time and their ability to thwart different proposed attacks. Table \ref{tab:final_comparison} shows this comparative analysis between different popular encryption techniques. We observe that random insertion of the key-gates \cite{roy2008epic} is the most primitive and simplest logic encryption approach, however, it offers no security against the state-of-the-art attacks. Fault Analysis (FA) based approaches \cite{rajendran2015fault} improve the output corruptibility at the cost of higher implementation complexity and longer encryption time compared to random logic encryption strategy. These methods also fail to prevent any of the proposed attacks. Strong Logic Locking (SLL) \cite{yasin2015improving} can prevent path sensitization \cite{yasin2015improving} and hill climbing \cite{plaza2015solving} attacks at the cost of lower output corruptibility, higher implementation complexity, and longer encryption time. Obfuscation Cell (OC) based encryption technique \cite{zhang2016practical} is an alternative of XOR/XNOR based encryption. However, a simple replacement of some wires of a netlist with the OCs cannot prevent any attack or produce high output corruption for wrong keys. To obtain either high output corruption or prevent some of the attacks, the OC-based method needs to be integrated with either FA or SLL based approach, respectively. Such integration nullifies the advantages of simplicity and quick encryption time of OC-based encryption technique. Logic cone prevention based technique \cite{lee2015improving} can prevent logic cone based attack, however, this method also needs to be integrated with SLL to prevent other attacks. External Key-Dependency based approach \cite{karmakar2017new} can prevent path sensitization, hill climbing and logic cone based attacks as well as offers high output corruption for wrong keys at the cost of four times hardware overhead compared to the conventional XOR/XNOR based logic encryption. SARLock method \cite{yasin2016sarlock}, integrated with SLL, can prevent path sensitization, hill climbing and SAT attacks at the cost of high hardware overhead and implementation complexity. However, this method offers low output corruptibility for wrong keys. Anti-SAT method \cite{xiemitigating}, integrated with FA based approach, can prevent SAT attack at the cost of hardware overhead and high implementation complexity. However, none of these approaches can prevent the scan-based attack. On the other hand, our proposed \textit{Encrypt Flip-Flop} strategy can prevent all the proposed attacks at the cost of low hardware overhead and medium implementation complexity. The encryption time of our proposed method is also very less. At the same time, our method also produces reasonable output corruption for wrong keys. All of the above observations justify the superiority of the \textit{Encrypt Flip-Flop} strategy over other state-of-the-art logic encryption techniques.

		\section{Conclusion and Future Works}
		In this paper, we have proposed a new scan based attack which can extract the keys of any logically encrypted circuit, irrespective of its size and the number of encryption keys, provided the circuit contains DfT infrastructure for infield testing and debugging. We have also proposed a new logic encryption strategy called \textit{Encrypt Flip-FLop}, which encrypts the outputs of selected flip-flops by inserting a MUX. The proposed strategy restricts the contrallability and observability of the scan chains of a circuit, thus, prevents scan based attack. It has inherent capability to thwart SAT attack, thus, unlike other SAT-preventive methods, \textit{Encrypt Flip-FLop} does not require extra hardware to develop SAT-resilience infrastructure. In contrary to other SAT-resilience methods, the proposed \textit{Encrypt Flip-Flop} strategy does not suffer from poor output corruptibility and threat of removal attack. The proposed low overhead encryption strategy can also prevent other state-of-the-art attacks. Simple design and low encryption times are the added advantages of the proposed encryption strategy. In the future work, we will focus on utilizing the proposed scan encryption technique to prevent the extraction of data encryption keys of the cryptographic ICs.

		
		%

		
		
		%

		\bibliographystyle{IEEEtran}
		\bibliography{ref.bib}
		\vspace{-35pt}
		\begin{IEEEbiography}[{\includegraphics[width=0.9in,height=1.1in,clip,keepaspectratio]{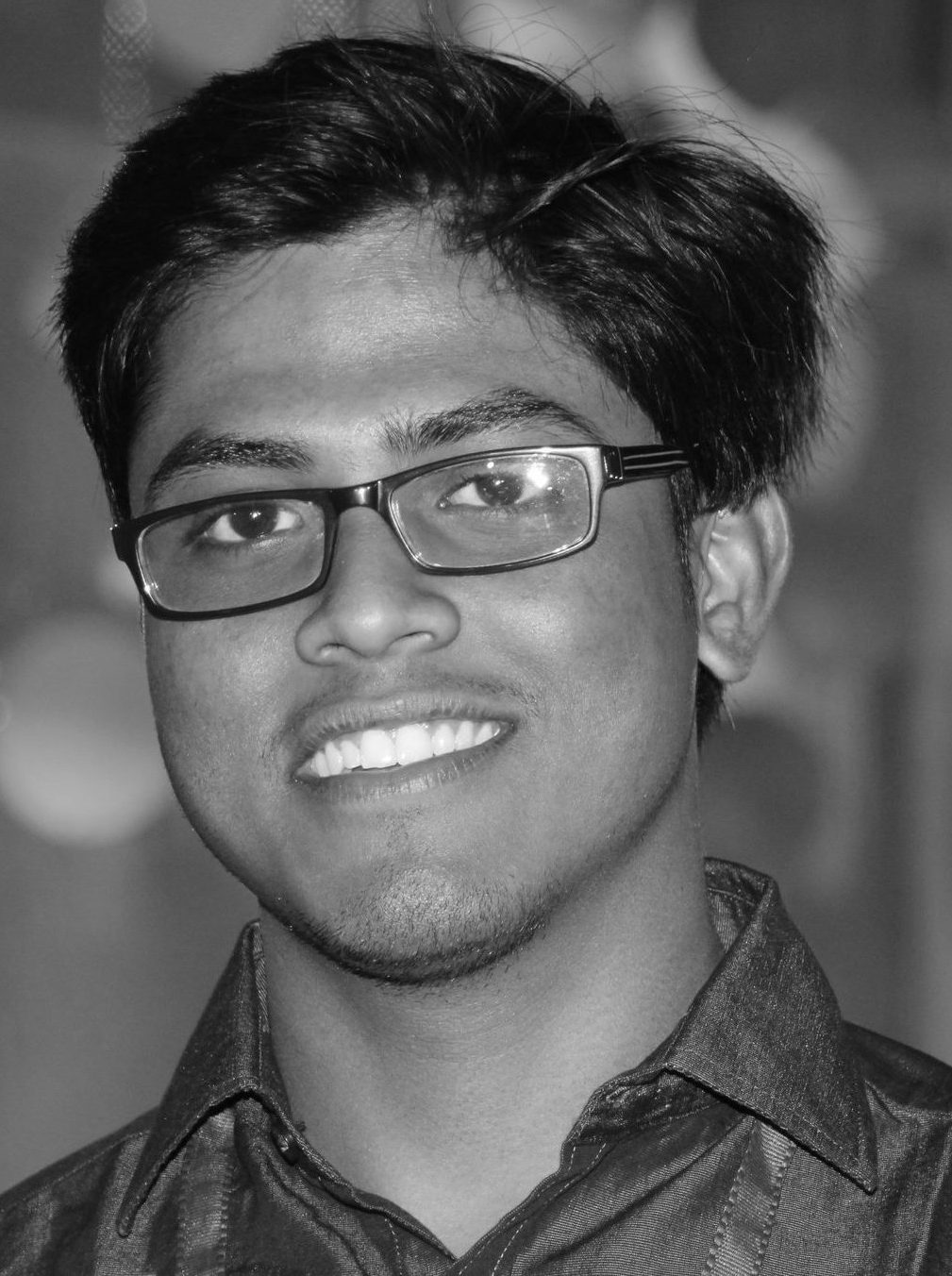}}]%
			{Rajit Karmakar}
			received his MS degree in Microelectronics and VLSI from Indian Institute of Technology, Kharagpur, India, in 2015. He is presently a PhD student in the Department of Electronics and Electrical Communication Engineering, Indian Institute of Technology, Kharagpur, India. His current research interests include hardware security and VLSI Testing. He is a student member of IEEE.
			\vspace{-20pt}
		\end{IEEEbiography}
		\vspace{-25pt}
		\begin{IEEEbiography}[{\includegraphics[width=1.3in,height=1.2in,clip,keepaspectratio]{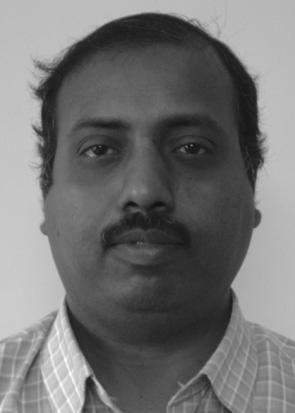}}]%
			{Santanu Chattopadhyay}
			received the B.E. degree in computer science and technology from Calcutta University, Kolkata, India, in 1990, the M.Tech. degree in computer and information technology and the PhD degree in computer science and engineering from the IIT Kharagpur, Kharagpur, India, in 1992 and 1996, respectively. He is a Professor with the Department of Electronics and Electrical Communication Engineering at IIT Kharagpur. He has published more than 150 technical papers in peer-reviewed journals and conferences. His current research interests include digital circuit design, testing and diagnosis, network-on-chip design and test, and low-power circuit design and test. He is a senior member of IEEE.
		\end{IEEEbiography}
		\vspace{-30pt}
		\begin{IEEEbiography}[{\includegraphics[width=1.3in,height=1.2in,clip,keepaspectratio]{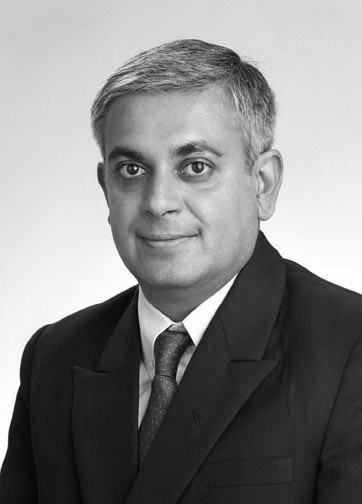}}]%
			{Rohit Kapur} is a Fellow in Synopsys. A recognized authorith in Test, he holds 25 patents and over a hundred publications. Rohit has made contribution to leading Synopsys products including DFTMAX, DFTMAX Ultra, and TetraMAX. He received his PhD in Computer Engineering from the University of Texas in 1992. He is a IEEE Fellow and TTTC 2nd Vice Chair. He was the group chair for the IEEE Core Test Language (CTL) standard. He is also in the editorial board of the Computer Magazine.
			
		\end{IEEEbiography}
	\end{document}